\documentclass[pteplogo]{ptephy}

\preprintnumber{XXXX-XXXX} 

\newcommand{\Vr}{{\cal R}}
\newcommand{\Xzero}{{\delta\! X}}

\begin{document}

\title{Gravitational wave echoes induced by a point mass plunging to a black hole}

\author{\name{Norichika~Sago}{1,2,*} and \name{Takahiro~Tanaka}{1,3},
}

\address{\affil{1}{Department of Physics, Kyoto University, Kyoto 606-8502, Japan}
\affil{2}{Advanced Mathematical Institute, Osaka City University, Osaka 558-8585, Japan}
\affil{3}{Center for Gravitational Physics, Yukawa Institute for Theoretical Physics, Kyoto University, Kyoto 606-8502, Japan}
\email{sago@tap.scphys.kyoto-u.ac.jp}}

\begin{abstract}%
Recently, the possibility of detecting gravitational wave echoes in the
data stream subsequent to the binary black hole mergers observed by LIGO
was suggested. Motivated by this suggestion, we presented templates
of echoes based on black hole perturbations in our previous work.
There, we assumed that the incident waves resulting in echoes are
similar to the ones that directly escape to the asymptotic infinity. 
In this work, to extract more reliable information  on the waveform of echoes 
without using the naive assumption on the incident waves, 
we investigate gravitational waves induced by a
point mass plunging into a Kerr black hole. We solve the linear
perturbation equation sourced by the plunging mass 
under the purely outgoing boundary condition at
infinity and a hypothetical reflection boundary condition near the horizon. 
We find that the low frequency component below the threshold of 
the super-radiant instability is highly suppressed, which is 
consistent with the incident waveform assumed in the previous analysis. 
We also find that the high frequency mode excitation is significantly 
larger than the one used in the previous analysis, 
if we adopt the perfectly reflective boundary condition independently of the frequency. 
When we use a simple template in which the same waveform as the 
direct emissions to infinity is repeated with the decreasing 
amplitude, the correlation between the expected signal and the template 
turns out to decrease very rapidly. 
\end{abstract}

\subjectindex{E01, E02, E31, E38}

\maketitle


\section{Introduction}
Since the direct detection of gravitational waves (GWs) was reported
by LIGO and Virgo collaboration~\cite{Abbott:2016blz, TheLIGOScientific:2016qqj},
the reported number of black hole merger events is rapidly increasing 
\cite{LIGOScientific:2018mvr, GWOSC-GWTC-1}. 
These data allow us to carefully examine the nature of black holes. 

The possible presence of gravitational wave echoes is one of the intriguing topics
stimulated by GW observations.
Abedi et al.~\cite{Abedi:2016hgu} analyzed the data succeeding to 
the BBH merger events observed by LIGO during
O1 observation run to search for signals of GW echoes, and claimed that
they found a tentative evidence for the echoes at false detection
probability of 0.011. Motivated by this work, several groups have done
the follow-up analyses
\cite{Ashton:2016xff, Abedi:2017isz, Westerweck:2017hus, Nielsen:2018lkf, Lo:2018sep, Uchikata:2019frs, Abedi:2020sgg}.

GW echoes after a compact binary coalescence (CBC) can be a 
probe of the exotic nature of black holes, since no GW echo 
is expected if the resulting object after coalescence is just 
an ordinary classical black hole. 
There is a possibility that GW echoes are induced, 
if the resulting object after merger is an exotic compact object (ECO) without horizon
\cite{Cardoso:2016rao, Cardoso:2016oxy},
e.g. gravastar \cite{Mazur:2004fk}, 
wormhole \cite{Visser:1995, Damour:2007ap},
firewall \cite{Almheiri:2012rt}, and so on
(Also see \cite{Cardoso:2019rvt}, which gives a comprehensive review on
ECOs and their tests).
Once GW echoes are observed, it would be possible to extract
the information about the ECO from the waveforms of the echoes.

Several works on the construction of the waveform of the GW echoes have
been done based on the black hole perturbation theory
\cite{Nakano:2017fvh, Mark:2017dnq, Burgess:2018pmm, Conklin:2019fcs,
Maggio:2019zyv, Micchi:2019yze, DAmico:2019dnn}.
In the case of an ordinary black hole spacetime, 
one can calculate GWs by solving the
perturbation equation ({\it e.g.} the Teukolsky equation in Kerr case) under
the pure-outgoing condition at infinity and the pure-ingoing one on the
horizon.
Even in the case of ECOs, we can consider the possibility that  
only a small neighborhood of the horizon might be modified, {\it i.e.}, 
the same spacetime as a black hole might be realized outside a 
hypothetical near-horizon boundary located slightly outside the horizon. 
Then, the perturbations in ECO spacetime can be 
expressed by the same equation as in general relativity, except for the modified boundary condition. 
The reflective boundary condition at the near-horizon boundary, 
instead of the pure-ingoing one, leads a series of GW echoes.  
As a result of this modified boundary condition, extra GWs
due to the reflection by the inner boundary is added to GWs 
observed at infinity, which can be calculated by multiplying 
the transfer function to the GWs that should have fallen into 
the horizon in the ordinary setup. 

The transfer function consists of the reflectance and the transmittance
determined by the effective potential of the perturbation equation, 
the position and reflectance of the inner boundary. The behavior of the transfer
function has been studied
well in previous research
\cite{Nakano:2017fvh, Mark:2017dnq, Burgess:2018pmm, Conklin:2019fcs,
Maggio:2019zyv, Micchi:2019yze, DAmico:2019dnn} 
by solving the scattering
problem in one-demension. By contrast, the waveform of the 
incident GWs, which is nothing but the ingoing waves absorbed by the black hole, 
has not been investigated extensively, especially in the Kerr case. In this paper 
we study this issue. 

In Sec. 2 after recapitulating the basic equations for the black hole 
perturbation theory based on the Sasaki-Nakamura equation, we make it 
clear how to compute the ingoing waveform. The variable that we use in 
the Sasaki-Nakamura equation is the one obtained by a transformation from 
the Teukolsky variable $\psi_4$. For this Teukolsky variable $\psi_4$, 
the amplitude of the ingoing waves are suppressed in the asymptotic regions 
at infinity and near the horizon, compared with the outgoing ones. 
Because of this, we cannot calculate the energy of the waves falling into the 
horizon directly from the asymptotic waveform of the Sasaki-Nakamura variable 
computed by using the standard Green function method. 
Here we give the explicit formula for the energy spectrum ingoing to the horizon.
Furthermore, 
based on this flux formula, we develop a method to impose a reflective 
boundary near the horizon, which applies even in the presence of the source term. 
In Sec. 3 we present the results of numerical calculation of echo waveform 
with the reflective boundary condition. We also discuss the detectability 
of the resulting waveform by using the previously proposed ways to generate 
the echo templates. 
Section 4 is dedicated for conclusion. 

\section{Basic equations}

\subsection{Plunge orbits}
The Kerr metric in the Boyer-Lindquist coordinates,
$(t, r, \theta, \varphi)$, is given by
\begin{eqnarray}
g_{\mu\nu}dx^\mu dx^\nu &=&
-\left(1-\frac{2Mr}{\Sigma}\right)dt^2
-\frac{4Mar\sin^2\!\theta}{\Sigma}dt\,d\varphi
+\frac{\Sigma}{\Delta}\,dr^2
\nonumber \\ &&
\hspace*{2cm} +\Sigma \,d\theta^2
+\left(r^2+a^2+\frac{2Ma^2r}{\Sigma}\sin^2\!\theta\right)
\sin^2\!\theta \,d\varphi^2, \label{eq:Kerr}
\end{eqnarray}
where $\Sigma=r^2+a^2\cos^2\theta$, $\Delta=r^2-2Mr+a^2$,
$M$ and $aM$ are the mass and the angular momentum of
the black hole, respectively.

In this work, we consider a point mass which is initially at rest
at infinity and falls into a black hole on the equatorial plane ($\theta=\pi/2$).
The equations of motion for the particle are given by
\begin{eqnarray}
r^2 \frac{dt}{d\tau} &=&
-a(a-L) + \frac{r^2+a^2}{\Delta(r)} P(r), \nonumber \\
r^2 \frac{dr}{d\tau} &=&
- \sqrt{\Vr(r)}, \nonumber \\
r^2 \frac{d\varphi}{d\tau} &=&
- (a - L) + \frac{a}{\Delta(r)}P(r),
\label{eq:EOM-plunge}
\end{eqnarray}
where $L$ is the angular momentum of the point mass,
$P(r)=r^2+a^2-aL$ and $\Vr(r)=2Mr^3-L^2r^2+2Mr(a-L)^2$.
(These equations correspond to the geodesic equations
with $E=1$ and $C=0$, where $E$ and $C$ are the specific
energy and Carter parameter of the point mass.)

\subsection{Sasaki-Nakamura equation}
The Sasaki-Nakamura equation is given by~\cite{Sasaki:1981kj}
\begin{equation}
\left( \frac{d^2}{dr_*^2} - F \frac{d}{dr_*} - U \right) X_{lm\omega}(r_*)
= S_{lm\omega}, \label{eq:SN-eq}
\end{equation}
where $r_*$ is defined by
\[
\frac{dr_*}{dr} = \frac{r^2+a^2}{\Delta}.
\]
The functions of $F$ and $U$ in Eq. (\ref{eq:SN-eq}) are given by
\begin{eqnarray}
F &=&
\frac{\Delta}{r^2+a^2} \frac{\gamma'}{\gamma}, \\
U &=&
\frac{\Delta U_0}{(r^2+a^2)^2} + G^2 + \frac{dG}{dr_*}
- \frac{\Delta G}{r^2+a^2} \frac{\gamma'}{\gamma},
\end{eqnarray}
where a prime means a differentiation with respect to $r$, and 
\begin{eqnarray*}
\alpha &=& 3iK' + \lambda + \frac{6\Delta}{r^2}
- \frac{iK}{\Delta^2}\beta, \\
\beta &=&
\Delta \left(
-2iK + \Delta' - \frac{4\Delta}{r}
\right), \\
\gamma &=&
\alpha \left( \alpha + \frac{\beta'}{\Delta} \right)
- \frac{\beta}{\Delta} \left(
  \alpha' + \frac{\beta}{\Delta^2} V \right), \\
G &=&
- \frac{\Delta'}{r^2+a^2} + \frac{r\Delta}{(r^2+a^2)^2}, \\
U_0 &=&
V + \frac{\Delta^2}{\beta} \left(
  \left( 2\alpha + \frac{\beta'}{\Delta} \right)'
  - \frac{\gamma'}{\gamma} \left(
  \alpha + \frac{\beta'}{\Delta} \right) \right), \\
V &=&
 -\frac{K^2+4i(r-M)K}{\Delta} + 8i \omega r + \lambda,\\
 K &=&(r^2+a^2)\omega -a m,
\end{eqnarray*}
with the eigenvalue of the spheroidal harmonics, $\lambda$.
(Here, we choose the free functions in the equations
given in Ref.~\cite{Sasaki:1981sx} as $f=h=1$ and $g=(r^2+a^2)/r^2$.)

The source term in the right hand side of Eq.(\ref{eq:SN-eq}) 
for a plunge orbit
in Eq.(\ref{eq:EOM-plunge}) is expressed by
\begin{equation}
S_{lm\omega} =
\frac{\gamma\Delta W_T}{(r^2+a^2)^{3/2}r^2}
\exp \left( -i \int^r \frac{K}{\Delta} dr \right),
\end{equation}
with
\begin{eqnarray}
W_T &=& W_{nn} + W_{n\bar{m}} + W_{\bar{m}\bar{m}}\,, \\
\frac{1}{\mu} W_{nn} &=&
f_0 \exp (i\chi)
+ \int_r^\infty dr_1 f_1 \exp(i\chi)
+ \int_r^\infty dr_1 \int_{r_1}^\infty dr_2 f_2 \exp(i\chi)\,, \\
\frac{1}{\mu} W_{n\bar{m}} &=&
g_0 \exp (i\chi)
+ \int_r^\infty dr_1 g_1 \exp(i\chi)\,, \\
\frac{1}{\mu} W_{\bar{m}\bar{m}} &=&
h_0 \exp(i\chi)
+ \int_r^\infty dr_1 h_1 \exp(i\chi)
+ \int_r^\infty dr_1 \int_{r_1}^\infty dr_2 h_2 \exp(i\chi)\,,
\end{eqnarray}
where
\begin{eqnarray}
f_0 &=&
 -\frac{1}{\omega} \frac{r^2\sqrt{\Vr}}{(r^2+a^2)^2} {\rm S}_c\,, \\
f_1 &=&
 \frac{f_0}{{\rm S}_c} \left[
 ( {\rm S}_1 + (a\omega-m) {\rm S}_0 ) \frac{ia}{r^2}
  + {\rm S}_c \left\{ \frac{2(a^2-r^2)}{r(r^2+a^2)}
     + \frac{\Vr'}{2\Vr} + i\eta
 \right\} \right]\,, \\
f_2 &=&
 \frac{i}{\omega} \frac{r^2\sqrt{\Vr}}{(r^2+a^2)\Delta}
 \left( 1 - \frac{P}{\sqrt{\Vr}} \right) \Biggl[
 \{ {\rm S}_1 + (a\omega-m){\rm S}_0 \} \frac{ia}{r^2}
\nonumber \\ &&
  + {\rm S}_c \left\{
    \frac{2a^2}{r(r^2+a^2)} + \frac{2r}{r^2+(L-a)^2}
    - \frac{(P+\sqrt{\Vr})'}{P+\sqrt{\Vr}} + i\eta
 \right\} \Biggr]\,, \\
\eta &=&
 \frac{(a\omega-m)(a-L)}{\sqrt{\Vr}}
 - \frac{am}{\Delta} \left( 1 - \frac{P}{\sqrt{\Vr}} \right)\,, \\
g_0 &=&
-\frac{a-L}{\omega} ( {\rm S}_1 + (a\omega-m) {\rm S}_0 )
 \frac{r^2}{r^2+a^2}\,, \\
g_1 &=&
g_0 \left[ \frac{2a^2}{r(r^2+a^2)} + i\eta \right]\,, \\
h_0 &=& -\frac{r^2 h_2}{2}, \quad
h_1 = -r h_2, \quad
h_2 = \frac{{\rm S}_0 (a-L)^2}{\sqrt{\Vr}}\,, \\
{\rm S}_0 &=& {}_{-2}{\rm S}_{lm}^{a\omega}(\pi/2), \quad
{\rm S}_1 = \frac{d}{d\theta} {}_{-2}{\rm S}_{lm}^{a\omega}(\pi/2)\,, \\
{\rm S}_c &=&
\left( a\omega - m - \frac{ia}{r} \right)
[ {\rm S}_1 + (a\omega-m) {\rm S}_0 ] - \frac{\lambda}{2} {\rm S}_0\,, \\
\chi &=&
\omega t - m \varphi + \int^r \frac{K}{\Delta} dr \,.
\end{eqnarray}
(These source terms are given in Ref.~\cite{Kojima:1984cj}
or Ref.~\cite{Nakamura:1987zz}.) 
${}_{-2}{\rm S}_{lm}^{a\omega}(\theta)$ is the spin-weighted
spheroidal harmonics with the spin weight $-2$.
Noticing that $\gamma=O(r^0)$ for $r\to\infty$ and $\gamma=O((r-r_+)^0)$ for $r\to r_+$, and $\Vr=O(r^{3})$ for $r\to\infty$ in the present setup,  
we find the asymptotic fall-off behaviors of the source term are given by 
\begin{eqnarray}
S_{lm\omega}=\left\{\begin{array}{ll}
    O\left(r^{-5/2}\right) ~~&\mbox{for}~ r\to\infty\,,\cr  \\
    O\left((r-r_+)\right) ~~&\mbox{for}~r\to r_+.  
\end{array} \right.\,
\label{eq:AympS}
 \end{eqnarray}
which correspond to $W_T=O(r^{1/2})$ and $W_T=O(1)$, respectively. 
The Sasaki-Nakamura variable $X_{lm\omega}(r_*)$ can be
converted to the Teukolsky variable $R_{lm\omega}(r_*)$ by the formula
\begin{equation}
R_{lm\omega}(r_*) = \Lambda[X_{lm\omega}(r_*)]
+ \frac{(r^2+a^2)^{3/2}}{\gamma} S_{lm\omega}, \label{eq:X2R}
\end{equation}
where the differential operator $\Lambda$ is defined by
\begin{equation}
\Lambda[X(r_*)] \equiv
\frac{1}{\gamma} \left[
\frac{(\alpha\Delta + \beta')}{\sqrt{r^2+a^2}} X(r_*)
- \frac{\beta}{\Delta} \frac{d}{dr}
   \left( \frac{\Delta}{\sqrt{r^2+a^2}} X(r_*) \right)
\right].
\end{equation}

Here, we should stress that the source term of 
the Sasaki-Nakamura equation is obtained by radially integrating the 
source term of the Teukolsky equation twice. Therefore, it 
should contain two arbitrary integration constants. 
The different choice of the source term $S_{lm\omega}$ leads to a 
different solution of $X_{lm\omega}$, but the resulting Teukolsky variable 
should be invariant. 
In fact, $R_{lm\omega}$ derived by using Eq.~(\ref{eq:X2R})
is invariant under the simultaneous transformations:
\begin{equation}
X_{lm\omega} \rightarrow
 \tilde{X}_{lm\omega} = X_{lm\omega} + \Xzero_{lm\omega}, \quad
S_{lm\omega} \rightarrow \tilde{S}_{lm\omega} 
= S_{lm\omega} + \delta S_{lm\omega}, \label{eq:X-trans}
\end{equation}
where
\begin{eqnarray}
\Xzero_{lm\omega} &=&
- \frac{\sqrt{r^2+a^2}}{r^4} \exp\left( -i\int \frac{K}{\Delta}dr \right)
\nonumber \\ && \times
\bigl[ ( 2r^2 - 2a^2 + \lambda r^2 - 4irK + 3ir^2K' ) c_0
\nonumber \\ && \hspace{5mm}
+ ( 6Mr - 6a^2 + \lambda r^2 - 6irK + 3ir^2K' ) c_1 r
\bigr]\,, \\
\delta S_{lm\omega} &=& \frac{\gamma\Delta ( c_0 + c_1 r ) }{(r^2+a^2)^{3/2}r^2}
\exp \left( -i \int^r \frac{K}{\Delta} dr \right).
\end{eqnarray}
In other words, $\Xzero_{lm\omega}$ and $\delta S_{lm\omega}$ satisfy the
following relation:
\begin{equation}
0 = \Lambda [\Xzero_{lm\omega}]
 + \frac{(r^2+a^2)^{3/2}}{\gamma} \delta S_{lm\omega}\,.
\label{eq:X0-dS-relation}
\end{equation}
One can also verify that the solution of the Sasaki-Nakamura equation
sourced by $\delta S_{lm\omega}$ is given by $\Xzero_{lm\omega}$. 

If we specify $c_0$ and $c_1$ as 
\begin{equation}
c_0 = -W_T(r_+) + r_+ W_T'(r_+)\,, \qquad
c_1 = -W_T'(r_+), \label{eq:coeff-dW}\,,
\end{equation}
we find that  
\begin{equation}
\tilde S_{lm\omega} =\left\{
    \begin{array}{ll}
          O(r^{-2}) ~~&\mbox{for}~r\to\infty\,,\\
          O((r-r_+)^3)~~&\mbox{for}~r\to r_+\,,\\
    \end{array}\right.
\end{equation}
which correspond to $W_T=O(r)$ for $r\to\infty$ and $W_T=O((r-r_+)^2)$
for $r\to r_+$, respectively.
Compared with the asymptotic behavior of the original source term~\eqref{eq:AympS}, 
the fall-off of the new source term $\tilde S_{lm\omega}$ is much faster 
near the horizon, while, as an expence to pay, 
the fall-off at infinity is slower. 

\subsection{Homogeneous solutions}
Let $X^{\infty,+}$ and $X^{\infty,-}$ stand for the homogeneous solutions of
the Sasaki-Nakamura equation~(\ref{eq:SN-eq})
that satisfy the purely outgoing $(+)$ and purely ingoing $(-)$ conditions
at infinity, respectively. 
We omit the index $lm\omega$ for simplicity, 
unless it is necessary. 
In the same manner, let $X^{H,\pm}$ denote the
purely outgoing and purely ingoing homogeneous solutions near the horizon.
The asymptotic forms are given by
\begin{eqnarray}
\begin{array}{ll}
X^{\infty,\pm} = e^{\pm i\omega r_*}\,, &
\mbox{for} \,\, r \to \infty \\
X^{H,\pm} = e^{\pm ikr_*} &
\mbox{for} \,\, r \to r_+\,.
\end{array}
\end{eqnarray}
In a similar manner, we define the homogeneous solutions of the
radial Teukolsky equation, $R^{\infty,\pm}$ and
$R^{H,\pm}$, which satisfy
\begin{eqnarray}
\begin{array}{ll}
R^{\infty,\pm} = r^{1 \pm 2} e^{\pm i\omega r_*}
 & \mbox{for} \,\, r \to \infty\,, \\
R^{H,\pm} = \Delta^{1 \mp 1} e^{\pm ikr_*}
 & \mbox{for} \,\, r \to r_+\,.
\end{array}
\end{eqnarray}

Substituting $X^{\infty,+}$ into Eq.~(\ref{eq:X2R}), 
we obtain the relation
\begin{equation}
\Lambda[X^{\infty,+}] =
\Gamma^{\infty,+} R^{\infty,+}\,,
\quad
\Gamma^{\infty,+} \equiv
\frac{4\omega^2}
{12i\omega M - \lambda(\lambda+2) + 12a\omega(a\omega-m)}\,.
\label{eq:Z2R_8p}
\end{equation}
When we apply this formula to an inhomogeneous solution, the source term in Eq.~(\ref{eq:X2R})
does not contribute independently of whether we use 
the original source term $S_{lm\omega}$ or the modified one 
$\tilde S_{lm\omega}$,
In the same way, the relation between $X^{H,+}$ and $R^{H,+}$ can be 
obtained as  
\begin{equation}
\Lambda[X^{H,+}] =
\Gamma^{H,+} R^{H,+},
\quad 
\Gamma^{H,+} \equiv
\frac{ -4 i \sqrt{2Mr_+} k (r_+ - M - 2ikMr_+ ) }
     {\gamma(r_+)}\,,
\end{equation}
where
\begin{eqnarray}
\gamma(r_+) &=&
-12\omega^2 r_+^2 + 8(9kM+i\lambda)\omega r_+ -96k^2 M^2
- 16ikM(\lambda+3)
\nonumber \\ &&
+ 12i\omega M + \lambda(\lambda+2)
+ \frac{48ikM^2}{r_+}.
\end{eqnarray}

To obtain the relation between $X^{H,-}$ and $R^{H,-}$ through Eq.~(\ref{eq:X2R}), 
the higher order corrections of $X^{H,-}$
with respect to $(r-r_+)$ are required, because the leading and the sub-leading 
order terms vanish in the expression for $R^{H,-}$.
The asymptotic form of the purely ingoing solution
is given up to $O((r-r_+)^2)$ by
\begin{equation}
X^{H,-} =
\left[ 1 + a_1 (r-r_+) + a_2 (r-r_+)^2 + O( (r-r_+)^3 ) \right] e^{-ikr_*},
\label{eq:XHin}
\end{equation}
where $a_1$ and $a_2$ are determined so that 
Eq.~(\ref{eq:SN-eq}) without the source term 
is satisfied at each order. 
Thus determined values of $a_1$ and $a_2$ are
\begin{eqnarray}
a_1 &=&
\frac{(r_+ - M) (r_+^2 + (\lambda-7)Mr_+ + 8M^2 ) - 2kMmar_+^2}
{ 2Mr_+ ( r_+ - M - 2ikMr_+ ) ( r_+ - M ) }\,, \label{eq:XHin-coeff1} \\
a_2 &=&
\frac{1}
{16M^2r_+^2 ( r_+ - M - ikMr_+ ) ( r_+ - M - 2ikMr_+ ) ( r_+ - M )^2 }
\nonumber \\
&& \times \Bigl\{
m^2 a^2 r_+^2
\left[ (4M^2k^2 + 2ikM - 1 ) r_+^2 - 2iM(Mk+1) r_+ - M^2 \right]
\nonumber \\ &&
+ 2maMr_+ \bigl[ 
 -6kr_+^4 - 2ir_+^3 ( 2 - ikM\lambda + 13ikM )
\nonumber \\ && \hspace{20mm}
 + iMr_+^2 ( 12 - 2M^2k^2 - 2ikM\lambda +35ikM )
\nonumber \\ && \hspace{20mm}
 - 3i (4 + 5ikM ) M^2 r_+ + 4iM^3 \bigr]
\nonumber \\ &&
- 2r_+^6 (1-7ikM) - 2Mr_+^5 ( 6 - 2ikM\lambda + 47ikM - 2\lambda )
\nonumber \\ &&
+ M^2 r_+^4 ( 156 -8ikM\lambda + \lambda^2 + 170ikM -26\lambda )
\nonumber \\ &&
- 2M^3 r_+^3 ( 244 - 2ikM\lambda - \lambda^2 + 25ikM - 28\lambda )
\nonumber \\ &&
+ M^4 r_+^2 ( 678 + \lambda^2 - 104ikM - 50\lambda )
\nonumber \\ &&
- 4M^5 r_+ ( 111 -16ikM - 4\lambda ) + 112M^6 
\Bigr\}\,.  \label{eq:XHin-coeff2}
\end{eqnarray}
Substituting Eq.~(\ref{eq:XHin}) into Eq.~(\ref{eq:X2R}) with Eqs.~(\ref{eq:XHin-coeff1}) and (\ref{eq:XHin-coeff2}), we obtain
\begin{equation}
\Lambda[X^{H,-}] =
\Gamma^{H,-} R^{H,-},
\quad
\Gamma^{H,-} \equiv
\frac{1}
{ 8\sqrt{2Mr_+} ( r_+ - M - 2ikMr_+ ) ( r_+ - M - ikMr_+ ) }.
\label{eq:Z2R_Hm}
\end{equation}
Here, one remark is in order 
when we apply this formula to an inhomogeneous solution. 
In general we cannot neglect the source term contribution in Eq.~(\ref{eq:X2R}). 
Neglecting the source term contribution can be justified only when 
we use the scheme in which the source term is suppressed near the horizon. 
Therefore, 
this relation can directly apply 
only when we consider 
the asymptotic behaviors of $\tilde X_{lm\omega}$, the solution 
obtained by considering the modified source term $\tilde S_{lm\omega}$. 

\subsection{Inhomogeneous solution with the ordinary boundary conditions}
Introducing a new variable $\xi$, defined by
$d\xi = \gamma dr_*$,
the Sasaki-Nakamura equation can be rewritten as
\begin{equation}
\left( \frac{d^2}{d\xi^2} - \frac{U}{\gamma^2} \right) X
= \frac{S}{\gamma^2}\,.
\end{equation}
We construct the retarded Green function for this equation 
by using the homogeneous solutions that 
satisfy appropriate boundary conditions, 
$X^{H,-}$ and $X^{\infty,+}$, as
\begin{equation}
G(\xi, \xi') = \frac{1}{W_G}
\left( X^{H,-}(\xi) X^{\infty,+}(\xi')
\theta(\xi'-\xi)
+ X^{\infty,+}(\xi) X^{H,-}(\xi')
\theta(\xi-\xi') \right)\,,
\label{eq:Green-fn}
\end{equation}
where $W_G$ is the Wronskian of $X^{\infty,+}$ and $X^{H,-}$,
defined by 
\[
W_G = W[X^{\infty,+}, X^{H,-}] =
X^{H,-}(\xi) \frac{d}{d\xi} X^{\infty,+}(\xi)
- X^{\infty,+}(\xi) \frac{d}{d\xi} X^{H,-}(\xi)\,,
\]
which is guaranteed to be constant in $\xi$. 
Using the Green function, we obtain the retarded solution
\begin{equation}
X^{S}(r_*) =
\frac{X^{\infty,+}(r_*)}{W_G}
\int_{-\infty}^{r_*} dr'_*
\frac{X^{H,-}(r'_*) S(r'_*)}{\gamma}
+ \frac{X^{H,-}(r_*)}{W_G} 
 \int_{r_*}^{\infty} dr'_*
\frac{X^{\infty,+}(r'_*) S(r'_*)}{\gamma}\, .
\label{eq:XS-sol}
\end{equation}
This solution $X^S$ satisfies the boundary condition that there is 
no incoming wave from the infinity, because $X^S$ does not contain 
such a component owing to the retarded nature that 
the above Green function possesses by construction, 
and the source term is suppressed in the limit $r\to \infty$. 
The boundary condition at $r\to r_+$ is also verified because of 
the retarded nature of the Green function and the fact that 
the source term does not contribute to the relation between $X$ and $R$ 
for the outgoing waves. 

The asymptotic form of Eq.~(\ref{eq:XS-sol}) at infinity is given by
\begin{equation}
X^{S} =
A^{\infty} X^{\infty,+} + O(r^{-5/2})\,,
\quad
A^{\infty} \equiv
\frac{1}{W_G} 
 \int_{-\infty}^{\infty} dr'_*
\frac{X^{H,-}(r'_*) S(r'_*)}{\gamma}\,,
\quad \text{for}\ r \to +\infty\,.
\label{eq:XS-sol8}
\end{equation}
Using the relation~(\ref{eq:Z2R_8p}), we translate
Eq.~(\ref{eq:XS-sol8}) into the Teukolsky variable at infinity as
\begin{eqnarray}
R
&=&
\Lambda[X^S]
+ \frac{(r^2+a^2)^{3/2}}{\gamma} S
\nonumber \\ &=&
Z^{\infty} R^{\infty,+} + O(r^{1/2}),
\quad
Z^{\infty} =
\Gamma^{\infty,+} A^{\infty}.
\end{eqnarray}

To convert Eq.~(\ref{eq:XS-sol}) into the Teukolsky variable $R$ 
near the horizon, we should make use of the invariance
of $R$ under the transformation of Eq.~(\ref{eq:X-trans})
with (\ref{eq:coeff-dW}).
Consider the solution for the transformed source, $\tilde{S}$,
through the Green's function as:
\begin{equation}
X^{\tilde{S}}(r_*) =
\frac{X^{\infty,+}(r_*)}{W_G} 
\int_{-\infty}^{r_*} dr'_*
\frac{X^{H,-}(r'_*) \tilde{S}(r'_*)}{\gamma}
+ \frac{X^{H,-}(r_*)}{W_G} 
 \int_{r_*}^{\infty} dr'_*
\frac{X^{\infty,+}(r'_*) \tilde{S}(r'_*)}{\gamma}.
\label{eq:tXS-sol}
\end{equation}
Now we consider the difference between $X^S$ and
$X^{\tilde{S}}$:
\begin{eqnarray}
X^{\tilde{S}} - X^{S}
&=&
\frac{X^{H,-}}{W_G} \int_{\xi}^\infty d\xi'
X^{\infty,+}(\xi') \delta S(\xi')
+ \frac{X^{\infty,+}}{W_G} \int_{-\infty}^\xi d\xi'
X^{H,-}(\xi') \delta S(\xi')
\nonumber \\ &=&
\frac{X^{H,-}}{W_G}
W[X^{\infty,+}, \Xzero]_{\xi}^\infty
+ \frac{X^{\infty,+}}{W_G}
W[X^{H,-}, \Xzero]_{-\infty}^\xi
\nonumber \\ &=&
\Xzero
+ \frac{W[X^{\infty,+}, \Xzero](\infty)}{W_G}
X^{H,-}
- \frac{W[X^{H,-}, \Xzero](-\infty)}{W_G}
X^{\infty,+},
\label{eq:dX_tSS}
\end{eqnarray}
where we use the following relation for an arbitrary homogeneous
solution, $X$:
\begin{eqnarray}
\int_{\xi_1}^{\xi_2} d\xi X(\xi) \delta S(\xi) &=&
\int_{\xi_1}^{\xi_2} d\xi X(\xi) \left(
\frac{d^2}{d\xi^2} - \frac{U}{\gamma^2}
\right) \Xzero(\xi)
\nonumber \\ &=&
\left[X(\xi) \frac{d}{d\xi} \Xzero
- \left( \frac{d}{d\xi}X(\xi) \right) \Xzero\right]_{\xi_1}^{\xi_2}
\nonumber \\ &&
+ \int_{\xi_1}^{\xi_2} d\xi \left[ \left(
\frac{d^2}{d\xi^2}
- \frac{U}{\gamma^2} \right) X(\xi)
\right] \Xzero(\xi)
\nonumber \\ &=&
\left[W[X(\xi), \Xzero(\xi)]\right]_{\xi_1}^{\xi_2}.
\end{eqnarray}
From Eqs.~(\ref{eq:X2R}) and (\ref{eq:dX_tSS}), we obtain
\begin{eqnarray}
R &=&
\Lambda[X^S] + \frac{(r^2+a^2)^{3/2}}{\gamma} S
\nonumber \\ &=&
\Lambda[X^{\tilde{S}} - \Xzero
 - w^{\infty,0}X^{H,-} + w^{H,0}X^{\infty,+} ]
 + \frac{(r^2+a^2)^{3/2}}{\gamma} S
\nonumber \\ &=&
\Lambda[X^{\tilde{S}}
 - w^{\infty,0}X^{H,-} + w^{H,0}X^{\infty,+} ]
 + \frac{(r^2+a^2)^{3/2}}{\gamma} \tilde{S}\,,
\label{eq:X2R_horizon}
\end{eqnarray}
where
\begin{equation}
w^{\infty,0} =
\frac{W[X^{\infty,+}, \Xzero](\infty)}{W_G},
\quad
w^{H,0} =
\frac{W[X^{H,-}, \Xzero](-\infty)}{W_G}.
\label{eq:normalized-Wron-80-H0}
\end{equation}
Since both $X^{H,-}$ and $\Xzero$ satisfy the
purely ingoing boundary condition at $r\to r_+$, we find $w^{H,0}=0$.
From this fact and $\tilde{S}=O((r-r_+)^3)$, we find the asymptotic
form of $R$ near the horizon is given by 
\begin{eqnarray}
R &=& 
Z^{H} R^{H,-} + O((r-r_+)^3),
\end{eqnarray}
with  
\begin{eqnarray}
Z^{H} &=&
\Gamma^{H,-} 
(\tilde{A}^{H} - w^{\infty,0}), \cr\cr
\tilde{A}^{H} &=&
\frac{1}{W_G} 
 \int_{-\infty}^{\infty} dr'_*
\frac{X^{\infty,+}(r'_*) \tilde{S}(r'_*)}{\gamma}.
\end{eqnarray}

\subsection{Solution with reflective boundary condition
near the horizon} \label{sec:solution_ref_BC}
We consider the case in which the ingoing wave to the black hole is reflected
by a hypothetical boundary near the horizon. 
To describe such a situation, we first consider the 
solution obtained by replacing the
pure-ingoing solution in the Green function, Eq.~(\ref{eq:Green-fn})
with the one that contains the reflection waves,
\begin{eqnarray}
X^{H,-} \rightarrow
X^{H, {\rm ref}} &\equiv&
X^{H, -} + {\rm R}_b \Phi X^{H,+}, \label{eq:XH_ref} \cr
&=& ( 1 - {\rm R}_b \Phi \hat{r} ) X^{H,-}
+ {\rm R}_b \Phi \hat{t} X^{\infty,+}\,,
\label{eq:newmode}
\end{eqnarray}
where ${\rm R}_b$ is the reflectance on the boundary defined by the 
square root of the ratio between 
the energies of the ingoing and outgoing waves,
and $\Phi$ is the factor that takes care of the non-trivial relation 
between the amplitude of the Sasaki-Nakamura variable and the energy spectrum.
The explicit formulae for $\Phi$ will be provided later.
In the second equality, we introduce $\hat{r}$ and $\hat{t}$ that satisfy the
relation $X^{H,+}+\hat{r}X^{H,-}=\hat{t} X^{\infty,+}$, which are the apparent reflection and 
transmission coefficients of the scattering problem 
for an incident outgoing wave from the horizon side reflected by the angular
momentum barrier.
Using the Green function with the above replacement~\eqref{eq:newmode}, 
we obtain a solution
\begin{equation}
\hat{X}^{S}(r_*) =
\frac{X^{\infty,+}(r_*)}{\hat{W}_G} 
\int_{r_{*b}}^{r_*} dr'_*
\frac{X^{H,{\rm ref}}(r'_*) S(r'_*)}{\gamma}
+ \frac{X^{H,{\rm ref}}(r_*)}{\hat{W}_G} 
 \int_{r_*}^{\infty} dr'_*
\frac{X^{\infty,+}(r'_*) S(r'_*)}{\gamma}\,,
\label{eq:newXG-sol}
\end{equation}
where $r_{*b}$ is the value of $r_*$ on the reflective boundary surface,  
and $\hat{W}_G$ is the Wronskian between $X^{\infty,+}$
and $X^{H,{\rm ref}}$,
\begin{equation}
\hat{W}_G = W[X^{\infty,+}(r_*), X^{H,{\rm ref}}(r'_*)]
= ( 1 - {\rm R}_b \Phi \hat{r} ) W_G\,.
\end{equation}
The coefficients $w^{\infty,0}$ and $w^{H,0}=0$ in the right hand side of 
Eq.(\ref{eq:X2R_horizon}) are now replaced with
\begin{eqnarray}
&&\hat w^{\infty,0} =
\frac{W[X^{\infty,+}, \Xzero](\infty)}{\hat W_G}
= \frac{w^{\infty,0}}{1-{\rm R}_b\Phi\hat{r}}
\,,
\cr
&&\hat{w}^{{\rm ref},0} =
\frac{W[X^{H,{\rm ref}}, \Xzero](r_{*b})}{\hat{W}_G}
=
\frac{{\rm R}_b \Phi}{1-{\rm R}_b \Phi \hat{r}}
\frac{W[X^{H,+}, \Xzero](r_{*b})}{W_G}\,. 
\end{eqnarray}
In contrast to the original case, $\hat{w}^{{\rm ref},0}$ 
does not vanish because of the presence of the outgoing
component in $X^{H,{\rm ref}}$. 
As a result, we obtain 
\begin{eqnarray}
R &=&
\Lambda[\hat X^{\tilde{S}}
 - \hat w^{\infty,0}X^{H,\rm{ref}} + \hat w^{\rm{ref},0}X^{\infty,+} ],
\label{eq:hatX2R_horizon}
\end{eqnarray}
as an expression for the Teukolsky variable valid near the horizon. 
The existence of the term with $X^{\infty,+}$, which contains the 
outgoing component near the horizon, in Eq.(\ref{eq:hatX2R_horizon}) 
is inconsistent with the boundary condition of $R$ that we impose 
near the horizon.
To make the requested retarded boundary conditions satisfied, 
we need to add a homogeneous solution 
to the solution~\eqref{eq:newXG-sol} as 
\begin{eqnarray}
\hat{X}^{{\rm mod}}(r_*) &=&
\hat{X}^{{S}}(r_*)- \hat{w}^{{\rm ref},0} X^{\infty,+}(r_*).
\label{eq:hatXG-sol}
\end{eqnarray}
The additional term proportional to $X^{\infty,+}(r_*)$ 
does not disturb the purely outgoing boundary condition at infinity. 
This modified solution behaves asymptotically at infinity like 
\begin{equation}
\hat{X}^{{\rm mod}} =
(\hat{A}^{\infty} - \hat{w}^{{\rm ref},0}) X^{\infty,+}
 + O(r^{-5/2})\,,
\quad \text{for}\ r \to +\infty\,,
\label{eq:hatXS-sol8}
\end{equation}
with
\begin{equation}
\hat{A}^{\infty} \equiv
\frac{1}{\hat{W}_G} 
 \int_{r_{*b}}^{\infty} dr'_*
\frac{X^{H,{\rm ref}}(r'_*) S(r'_*)}{\gamma}
\approx
\frac{1}{\hat{W}_G} 
 \int_{-\infty}^{\infty} dr'_*
\frac{X^{H,{\rm ref}}(r'_*) S(r'_*)}{\gamma}\,,
\end{equation}
where in the last approximate equality 
we neglect the error due to the change of the integration range. 
The asymptotic amplitude $\hat{A}^{\infty}$ in the current problem 
can be expressed in terms of the
amplitudes $A^\infty$ and $\tilde A^H$ evaluated in the original case
with the ordinary boundary conditions as
\begin{eqnarray}
\hat{A}^{\infty} &=&
\frac{1}{(1-{\rm R}_b \Phi \hat{r}) W_G} 
 \int_{-\infty}^{\infty} dr'_*
\frac{(1-{\rm R}_b \Phi \hat{r})X^{H,-}(r'_*)
 + {\rm R}_b \Phi \hat{t} X^{\infty, +}(r'_*)}{\gamma} S(r'_*)
\nonumber \\ &=&
A^{\infty}
+ \frac{{\rm R}_b \Phi \hat{t}}{(1-{\rm R}_b \Phi \hat{r})W_G}
 \int_{-\infty}^{\infty} dr'_*
\frac{X^{\infty,+}(r'_*)
\left\{ \tilde{S}(r'_*) - \delta S(r'_*) \right\}}{\gamma}
\nonumber \\ &=&
A^{\infty}
+ \frac{{\rm R}_b \Phi \hat{t}}{1-{\rm R}_b \Phi \hat{r}}\left[ \tilde{A}^{H}
  - \frac{W[X^{\infty,+}, \Xzero](\infty)}{W_G}
  + \frac{W[X^{\infty,+}, \Xzero](-\infty)}{W_G}  \right]
\nonumber \\ &=&
A^{\infty}
+ \frac{{\rm R}_b \Phi \hat{t}}{1-{\rm R}_b \Phi \hat{r}}
\left[ \tilde{A}^{H} - w^{\infty,0}
+ \frac{\hat{r}}{\hat{t}} w^{H,0} 
+ \frac{W[X^{H,+}, \Xzero](-\infty)}{\hat{t}W_G}  \right]
\nonumber \\ &=&
A^{\infty}
+ \frac{{\rm R}_b \Phi \hat{t}}{1-{\rm R}_b \Phi \hat{r}}
\left( \tilde{A}^{H} - w^{\infty,0} \right)
+ \hat{w}^{{\rm ref},0}\,.
\end{eqnarray}
Using the relation of Eq.~(\ref{eq:Z2R_8p}), we translate
Eq.~(\ref{eq:hatXS-sol8}) into the Teukolsky variable in the limit $r\to\infty$ as
\begin{eqnarray}
\hat{R}
&=&
\Lambda[(\hat{A}^{\infty} - \hat{w}^{{\rm ref},0})
 X^{\infty,+}+O(r^{-5/2})]
+ \frac{(r^2+a^2)^{3/2}}{\gamma} S_X
\nonumber \\ &=&
\hat{Z}^{\infty} R^{\infty,+} + O(r^{1/2})\,,
\end{eqnarray}
with
\begin{eqnarray}
\hat{Z}^{\infty} &=&
Z^{\infty} + {{\cal K}} Z^{H}, \quad
{{\cal K}} =
 \frac{{\rm R}_b \Phi \hat{t}}{1-{\rm R}_b \Phi \hat{r}}
   \frac{\Gamma^{\infty,+}}{\Gamma^{H,-}}\,. \label{eq:hatZ8_Z8H}
\end{eqnarray}
The relation~(\ref{eq:hatZ8_Z8H}) is the gravitational wave counterpart 
of Eq.~(2.25) in Ref.~\cite{Mark:2017dnq}, and ${{\cal K}}$ shown
here is the transfer function.

\subsection{Reflectance on the boundary surface}
Here, we determine $\Phi$ in the expression for $X^{H,{\rm ref}}$ given by \eqref{eq:newmode}. 
The corresponding Teukolsky radial function is derived as
\begin{equation}
R^{H,{\rm ref}} = \Lambda\left[ X^{H,{\rm ref}} \right]
=
\Gamma^{H,-} R^{H,-}
+ \Gamma^{H,+} {\rm R}_b\Phi R^{H,+}.
\end{equation}
Here, we quote the formulae for the energy spectra of the ingoing and outgoing waves 
across the boundary surface given by \cite{Teukolsky:1974yv, Nakano:2017fvh}
\begin{equation}
\left( \frac{dE}{d\omega} \right)^{H,-}
=
\mu^2 \frac{\epsilon_{H,-}^2}{4\pi\omega^2}
\left| \Gamma^{H,-} \right|^2,
\quad
\left( \frac{dE}{d\omega} \right)^{H,+}
=
\mu^2 \frac{\epsilon_{H,+}^2}{4\pi\omega^2}
\left| \Gamma^{H,+} {\rm R}_b\Phi \right|,
\end{equation}
with
\begin{eqnarray}
\epsilon_{H,-}^2 &=&
\frac{256(2Mr_+)^5 (k^2+4\epsilon^2) (k^2+16\epsilon^2) k \omega^3}
{|C_{SC}|^2}, \\
\epsilon_{H,+}^2 &=&
\frac{\omega^3}{k(2Mr_+)^3(k^2+4\epsilon^2)}, \\
\epsilon &=& \frac{\sqrt{M^2-a^2}}{4Mr_+}, \\
|C_{SC}|^2 &=&
\left[ (\lambda+2)^2 + 4a\omega m - 4a^2\omega^2 \right]
\left[ \lambda^2 + 36a\omega m - 36a^2\omega^2 \right]
\nonumber \\ &&
+ 48 a\omega (2\lambda + 3) ( 2a\omega - m )
+ 144 \omega^2 ( M^2  - a^2 ).
\end{eqnarray}
Since the reflectance on the boundary ${\rm R}_b$ should be 
defined to satisfy
\begin{eqnarray}
|{\rm R}_b|^2 =\left( \frac{dE}{d\omega} \right)^{H,+}
\bigg/
\left( \frac{dE}{d\omega} \right)^{H,-}
&=&
\left| \frac{\epsilon_{H,+} \Gamma^{H,+} {\rm R}_b \Phi}
{\epsilon_{H,-} \Gamma^{H,-} } \right|^2\,,
\end{eqnarray}
we obtain 
\begin{equation}
\Phi =
\frac{\epsilon_{{\rm H},-} \Gamma^{{\rm H},-}}
{\epsilon_{{\rm H},+} \Gamma^{{\rm H},+}}
e^{-2ikr_{*b}}\,.
\end{equation}
For later use, we also introduce 
the phase shift on the boundary given by
\begin{eqnarray}
\Delta\phi 
\equiv
\text{arg} ({\rm R}_b)
=
\text{arg}\left(
\frac{\Gamma^{H,+} A^{{\rm ref}} e^{ikr_{*b}}}
{\Gamma^{H,-} e^{-ikr_{*b}}}
\right)\,.
\end{eqnarray}
Since ${\rm R}_b$ depends on the unknown property of the inner boundary, 
it cannot be specified uniquely. 
Later, we discuss a few simple models of ${\rm R}_b$. 

\subsection{Reflectance and transmittance on the angular momentum barrier}
In Sec.~\ref{sec:solution_ref_BC}, we introduce $\hat{r}$ and $\hat{t}$ that satisfy
$X^{H,+}+\hat{r}X^{H,-}=\hat{t}X^{\infty,+}$. These coefficients are determined by 
the property of the scattering due to the angular momentum barrier. 
By using the current notation, 
the reflectance and the phase shift given in Ref.~\cite{Nakano:2017fvh}
are represented by 
\begin{equation}
{\rm R} =
 \left| \frac{\epsilon_{H,-} \Gamma^{H,-}}
{\epsilon_{H,+} \Gamma^{H,+}} \right|^2 |\hat{r}|^2,
\quad
\phi_{-2}(f) =
 \text{arg} \left( \frac{\hat{r}\Gamma^{H,-}}{\Gamma^{H,+}} \right).
\end{equation}
$\hat{r}$ and $\hat{t}$ can be calculated from the Wronskians between the
homogeneous solutions as
\begin{equation}
\hat{r} = -\frac{W[X^{\infty,+}, X^{H,+}]}{W_G},
\quad
\hat{t} = \frac{W[X^{H,+}, X^{H,-}]}{W_G}.
\end{equation}

\section{Results}
Based on the formulation shown in the previous section, we compute
the gravitational waves produced by a point particle which is initially
at rest at infinity and falls on the equatorial plane to a black hole.
We take $M=1$ for all computation in this paper.

\subsection{Energy spectrum and transfer function}
\begin{figure}[!h]
\centering\includegraphics[width=8cm]{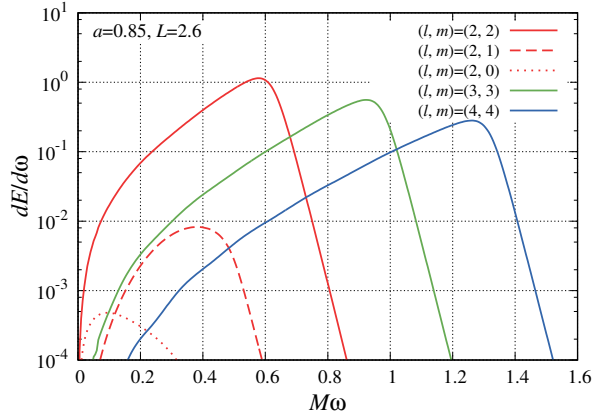}
\caption{Energy spectrum radiated to infinity in the case of a
corotating orbit with $a=0.85M$ and $L=2.6M$.}
\label{fig:dEdt8_a085_Lz26}
\end{figure}

To check our numerical code, we first compute the energy of GWs 
radiated to infinity by considering 
an infalling particle with the ordinary boundary conditions.
In Fig.\ref{fig:dEdt8_a085_Lz26}, we show the energy spectra of
$(l,m)=(2,2), (2,1), (2,0), (3,3), (4,4)$ in the case of $a=0.85M$ and
$L=2.6M$. This plot is consistent with that in Fig.3(a) of
\cite{Kojima:1984cj}, except for the difference of the factor $2$,
which comes from the difference in the definition of the 
energy spectrum
$(dE/d\omega)$: the spectrum in \cite{Kojima:1984cj} corresponds to
the one-side spectrum density, while ours is the two-side one.

\begin{figure}[!h]
\centering\includegraphics[width=7cm]{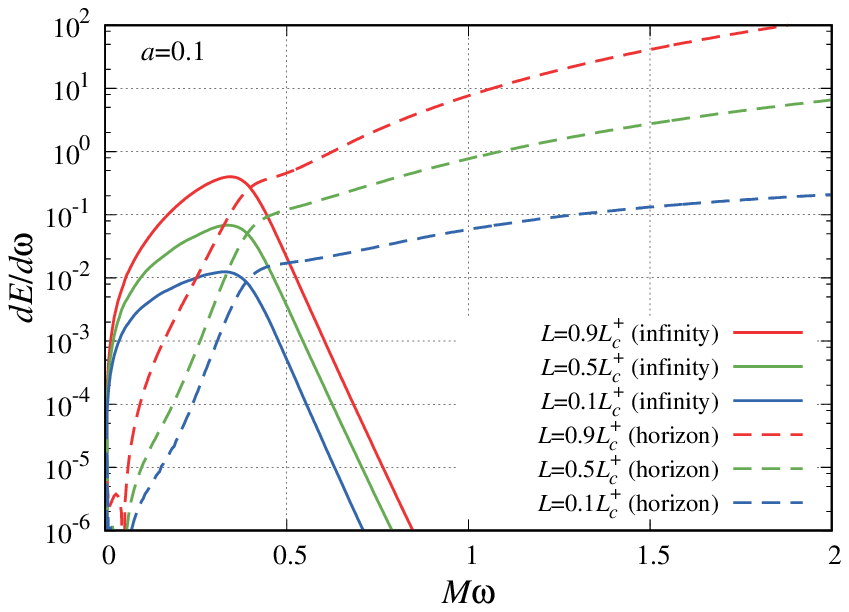}
\centering\includegraphics[width=7cm]{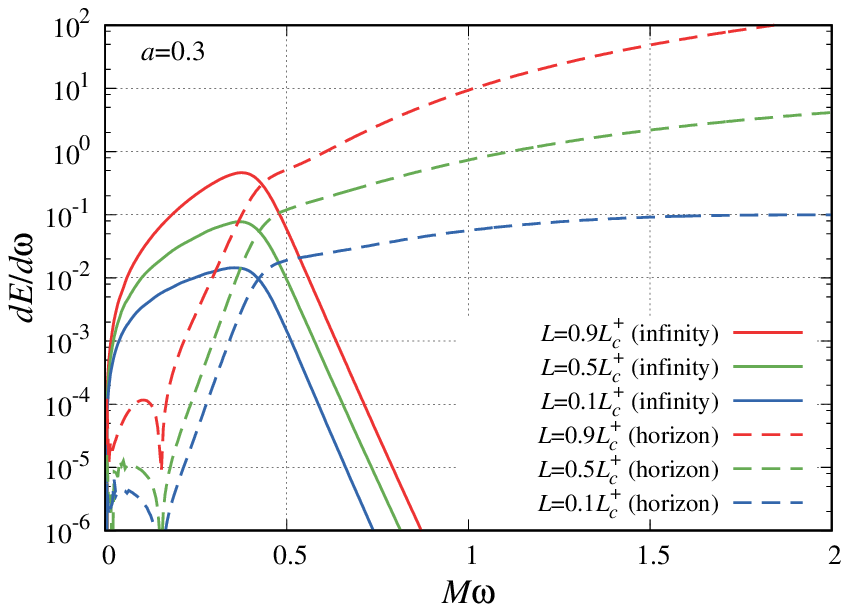} \\
\centering\includegraphics[width=7cm]{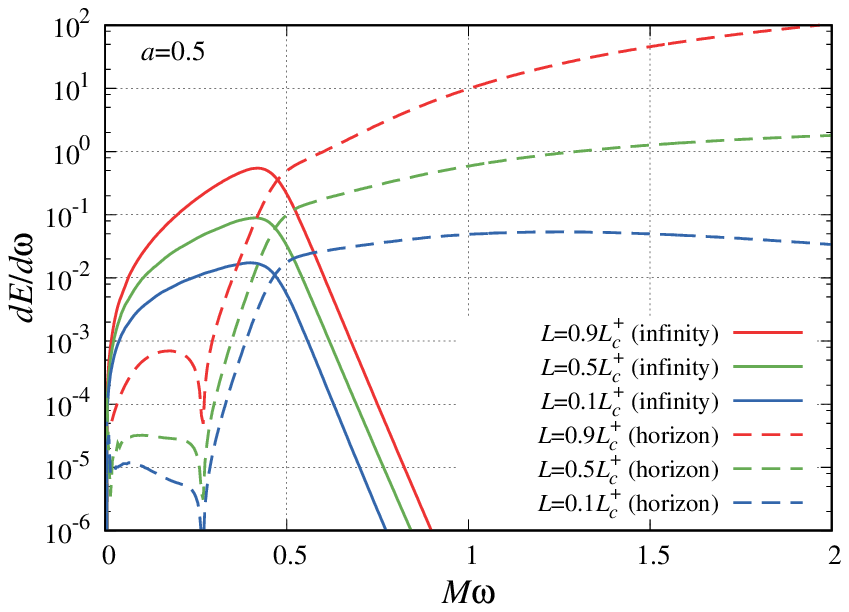}
\centering\includegraphics[width=7cm]{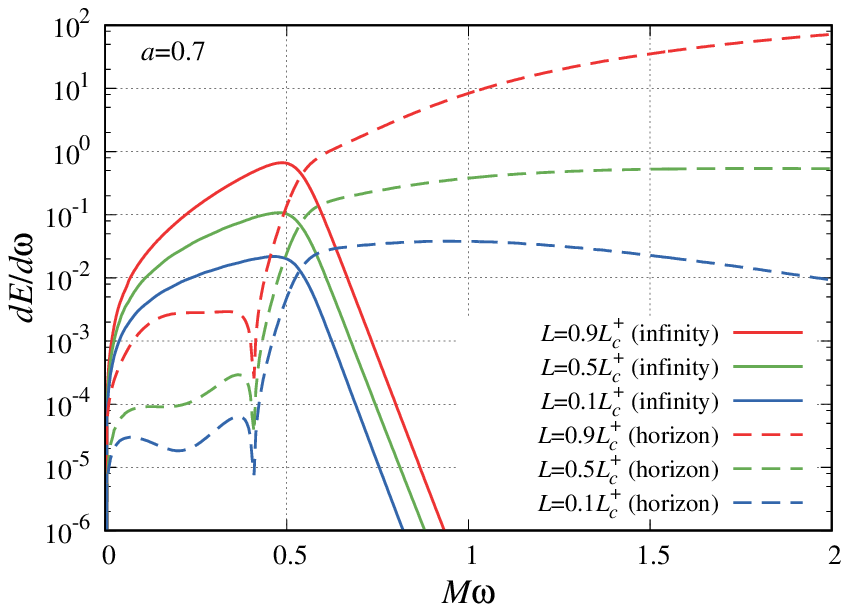}
\caption{Energy spectra of GWs radiated to infinity and to the horizon of
$(l, m)=(2,2)$ for several sets of parameters, $(a, L)$.
We choose four values of the Kerr parameter, $a=0.1$ (top left),
$a=0.3$ (top right), $a=0.5$ (bottom left), and $a=0.7$ 
(bottom right). The solid and dashed curves in each panel show the 
energy spectra to infinity and to the horizon, respectively.
Three lines correspond to different values of the angular momentum,
$L=0.9L_c^+$ (red), $L=0.5L_c^+$ (green), and $L=0.1L_c^+$ (blue) from
top to bottom.}
\label{fig:dEdt_BH}
\end{figure}

Next we compute the energies both radiated to the
infinity and absorbed by the black hole with ordinary boundary
conditions for several sets of the parameters $(a,L)$.
In Fig.\ref{fig:dEdt_BH}, we show both the energy spectra
of the $(l,m)=(2,2)$ mode for
$a=\{0.1, 0.3, 0.5, 0.7\}M$ and $L=\{0.1, 0.5, 0.9\}L_c^+$,
where $L_c^\pm$ is the critical value of the angular momentum
\[
L_c^\pm = \pm 2M ( 1 + \sqrt{1\mp a/M} ).
\]
In the above equation, the upper sign corresponds to corotating
orbits and the lower to counterrotating ones. Our focus is on plunge orbits, 
and hence we only discuss $L$ that satisfies $L_c^- < L < L_c^+$.

The energy spectra of GWs absorbed by the horizon is largely suppressed 
at low frequencies. This result turns out to be consistent with 
the feature of the waveform model proposed in our previous work 
\cite{Uchikata:2019frs,Nakano:2017fvh}. However, 
the spectra emitted to infinity and to the horizon are not similar 
at all. 
The suppression at low frequencies suggested in Ref.\cite{Nakano:2017fvh} 
is the one caused by the 
small transmission coefficient due to the angular momentum barrier, 
while in the present model the original amplitude of GWs to be 
reflected by the hypothetical boundary near the horizon is already 
suppressed at low frequencies. 
This lack of similarity 
means that it cannot be an optimal choice to use the waveform 
of GWs directly emitted to infinity as the seed to generate 
the echo template. 

We can also find that the energy spectrum to the horizon 
is dominated by the higher frequency band than the quasi-normal 
mode (QNM) frequency.
Here, we consider GWs excited by an in-falling point mass. 
If we replace the source with a finite size body, there might arise some suppression at high frequencies for ingoing waves falling into 
the black hole. 
To suppress the influence of the high frequency modes, 
which might be an artifact of using the point particle model,
it might be appropriate to reduce the amplitude of the 
high frequency component contained in the waves reflected by the hypothetical boundary near the horizon. 

There is also another motivation to introduce a cutoff to 
the high frequency modes, related to the expected property of 
the hypothetical reflective boundary near the horizon. 
There are some proposals suggesting that only low frequency gravitational waves are reflected back by the boundary~\cite{Bekenstein:1995ju,Wang:2019rcf, Oshita:2019sat}. 
In Ref.~\cite{Bekenstein:1995ju} actually proposed was 
the discretization of the horizon area, from which we naively 
expect the avoidance of the absorption of low frequency GWs since 
the state after the absorption is absent.  
References~\cite{Wang:2019rcf, Oshita:2019sat} 
proposed very different independent arguments that suggest the reflective  
boundary selective to low frequency GWs. 
To take into account such a possibly 
expected property of the boundary, here 
we consider a few simple models that give a frequency dependence to the reflectance of the boundary near the horizon. 
The simplest one is the sharp cut-off model, given by 
\begin{equation}
{\rm R}_b =
\left\{ \begin{array}{ll}
    1 & (|\omega|<\omega_c) \\
    0 & \mbox{(otherwise)}
  \end{array} \right., \label{eq:Rb-cutoff}
\end{equation}
where $\omega_c>0$ is a parameter introduced as the cutoff frequency.
We also introduce a model with the reflectance, 
\begin{equation}
{\rm R}_b =
\exp \left[ - \left( 
\frac{\omega-\omega_{{\rm QNM}}}{\sigma_\omega}
\right)^2 \right], \label{eq:Rb-Gaussian-model}
\end{equation}
where $\omega_{{\rm QNM}}$ is the frequency of the least damped
quasi-normal mode for the $m=2$ mode in Kerr spacetime
(here we use the fitting function given in\cite{Berti:2005ys}),
and $\sigma_\omega$ is a free parameter.
We refer to this model as the Gauusian model. 
In addition, we consider a reflectance model 
proposed in Ref.~\cite{Wang:2019rcf, Oshita:2019sat},
\begin{equation}
{\rm R}_b=e^{-|k|/T_H}, \label{eq:Rb-Q-model}
\end{equation}
where $T_H:=(r_+^2-a^2)/(4\pi r_+ (r_+^2+a^2))$ 
is the Hawking temperature. We refer to this model as the 
quantum black hole (QBH) model. 

\begin{figure}[!ht]
\centering\includegraphics[width=7cm]{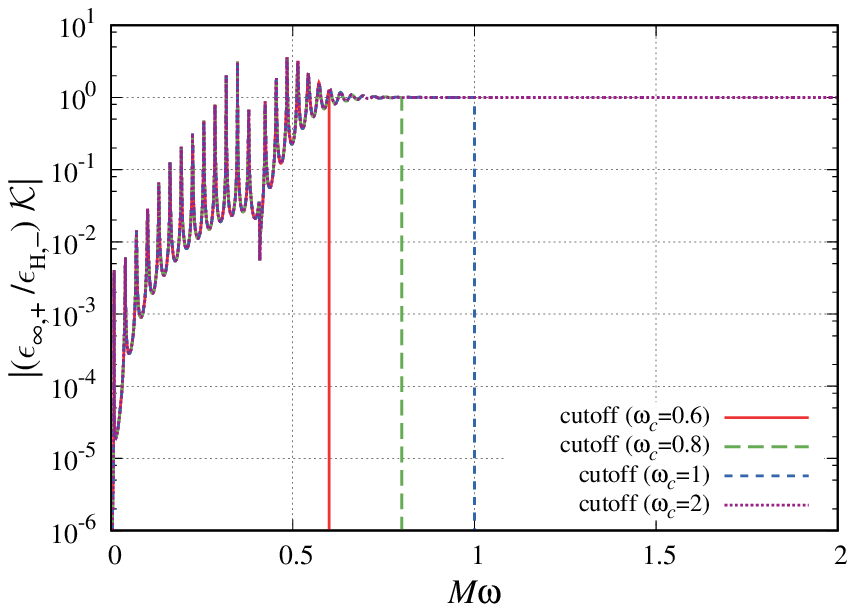}
\centering\includegraphics[width=7cm]{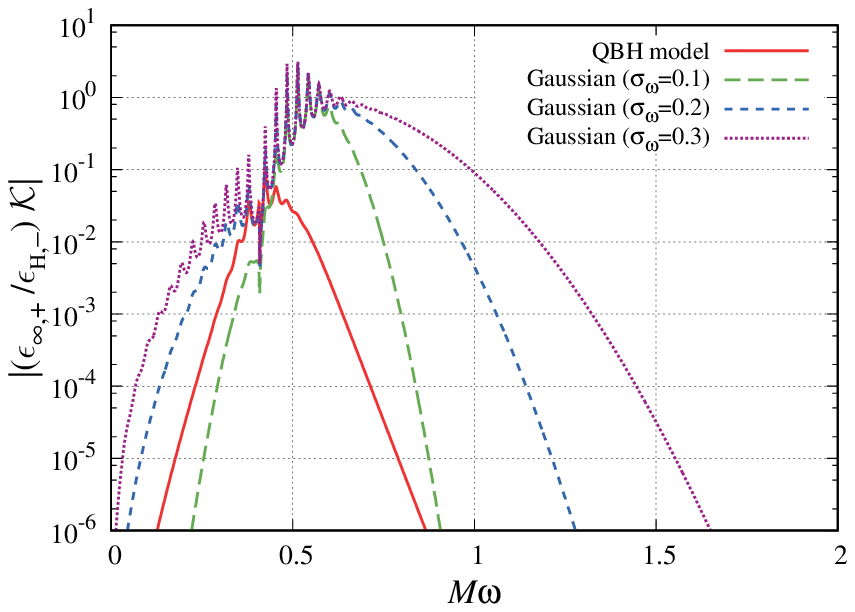}
\caption{Transfer functions with the reflective surface near
the horizon, $(\epsilon_{\infty,+}/\epsilon_{H,-})|\cal K|$. 
Here we plot the cases with the reflectance of
Eq.~(\ref{eq:Rb-cutoff}) with the cutoff parameter
$\omega_c={0.6, 0.8, 1, 2}$ on the left panel, while 
the cases with those of 
Eqs.~(\ref{eq:Rb-Gaussian-model}) with
$\sigma_\omega=(0.1, 0.2, 0.3)$ and (\ref{eq:Rb-Q-model})
on the right panel.
We fix $(l,m)=(2,2)$,
$a=0.7$ and $r_{*b}=-100M$ for all cases.}
\label{fig:transfer-fun}
\end{figure}

In Fig.\ref{fig:transfer-fun}, we show the transfer function 
$|(\epsilon_{\infty,+}/\epsilon_{H,-})\cal K|$ 
with the various choices of the reflectance on the inner boundary,
for $(l,m)=(2,2)$ and $a=0.7$. The factor $|(\epsilon_{\infty,+}/\epsilon_{H,-})|$ is multiplied so that the plot 
shows the square root of 
the ratio of the energy flux that reaches the infinity 
compared with that falls into the black hole in the case 
without the reflective boundary. 
We fix the position of the boundary surface at $r_{*b}=-100M$
for all cases.

Once we obtain 
$Z^H$ under the ordinary boundary
conditions and transfer function ${\cal K}$, we can construct
the echo amplitude and its waveform 
measured at infinity under the presence of the 
reflective boundary near the horizon, through Eq.~(\ref{eq:hatZ8_Z8H}).

\begin{figure}[!h]
\centering\includegraphics[width=10cm]{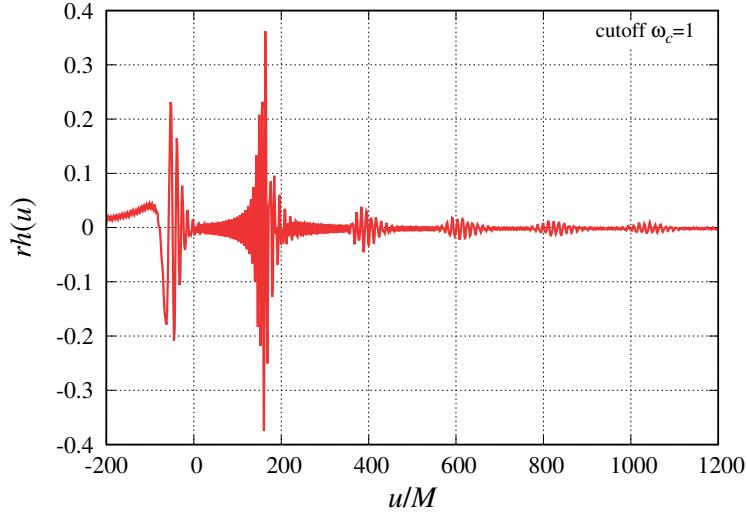} \\
\caption{Waveform of the $(l,m)=(2,2)$ mode for the cutoff model
with $\omega_c=1$.
Here we set $(a,L)=(0.7,0.9L_c^+)$ and $r_{*b}=-100M$.
}
\label{fig:waveform}
\end{figure}

\subsection{Comparison with the waveform constructed from the outgoing wave}
The time domain waveform observed at infinity can be constructed
from the asymptotic form of the Teukolsky variable,
$R = Z r^3 e^{i\omega r_*}$ by
\begin{equation}
h(u) =
\int d\omega e^{-i\omega u} H(\omega;Z), \quad
H(\omega;Z) =
-\frac{2}{r}\sum_{lm} \frac{Z}{\omega^2}
\frac{e^{im\varphi}}{\sqrt{2\pi}} {\rm S}(\theta)
\label{eq:TD-waveform}
\end{equation}
with $u=t-r_*$.
From Eq.~(\ref{eq:hatZ8_Z8H}), the waveform in the reflective
boundary case, $\hat{h}^\infty(u)$ is given by
\begin{equation}
    \hat{h}^\infty(u) = h^\infty(u) + h_{{\rm echo}}(u), 
\end{equation}
where 
\begin{eqnarray}
h^\infty(u) &=&
\int d\omega e^{-i\omega u}
H(\omega;Z^\infty)\,,
\label{eq:waveform-BHcase}
\end{eqnarray}
correspond to
the waveform observed at infinity in the ordinary BH boundary
case, and
\begin{eqnarray}
h_{{\rm echo}}(u) &=&
\int d\omega e^{-i\omega u} H(\omega;{\cal K}Z^H), 
\label{eq:waveform-echoes}
\end{eqnarray}
is that of the series of echoes in the reflective boundary
case.
As a demonstration, we give the time-domain waveform of the
$(l,m)=(2,2)$ mode for the cutoff model with $\omega_c=1$ in
Fig.~\ref{fig:waveform}. 
Even with the high-frequency cutoff, one 
can recognize that the first echo is very loud. 

In our previous work, we constructed the waveform of echoes using the
merger-ringdown waveform observed at infinity as the seed, 
instead of the waveform of GWs falling into the black hole. 
Namely, in the present context 
our previous waveform would correspond to the one obtained by 
replacing $Z^H$ in Eq.~(\ref{eq:waveform-echoes}) with $Z^\infty$,
\begin{equation}
\bar{h}_{{\rm echo}}(u) =
\int d\omega e^{-i\omega u} H(\omega;{\cal K}Z^\infty)\,.
\label{eq:waveform-echoes-dash}
\end{equation}

To compare $h_{{\rm echo}}$ and $\bar{h}_{{\rm echo}}$,
we introduce the overlap between two waveforms by
\begin{eqnarray}
\rho &=& \max_{\Delta t, \Delta\phi, \bar{r}_{*0}}
\frac{(h_{{\rm echo}}|\bar{h}_{{\rm echo}})}
{\sqrt{(h_{{\rm echo}}|h_{{\rm echo}})}
 \sqrt{(\bar{h}_{{\rm echo}}|\bar{h}_{{\rm echo}})}},
 \label{eq:define_rho} 
\end{eqnarray}
with
\begin{eqnarray}
(f|g) &\equiv&
\frac{1}{2\pi} \int_{-\infty}^{\infty}
\left[ F(\omega) G^*(\omega) 
+ F^*(\omega) G(\omega) \right] d\omega, \nonumber
\end{eqnarray}
where $\Delta t$, $\Delta\phi$ are the shifts of time and phase
between the two waveforms, and $\bar{r}_{*b}$, which is the location 
of the reflective boundary surface for $\bar{h}_{{\rm echo}}$, controls the time interval between neighboring echoes. 
Here, we assume white noise spectrum, which would be a good approximation 
as long as we are interested in the waveform 
whose power is localized in a narrow frequency band. 
In the above equation, we
marginalize $\Delta t$, $\Delta\phi$ and $\bar{r}_{*b}$ in
$\bar{h}_{{\rm echo}}$ to maximize $\rho$, while we fix the other
parameters to the same values as $h_{{\rm echo}}$.
We also define a new estimator for the detectability of the signal 
when we use a specified echo template, {\it effective filtered amplitude}
(EFA), by
\begin{equation}
{\rm EFA} = \max_{\Delta t, \Delta\phi, \bar{r}_{*b}}
\frac{(h_{{\rm echo}}|\bar{h}_{{\rm echo}})}
{\sqrt{(h^\infty|h^\infty)}
 \sqrt{(\bar{h}_{{\rm echo}}|\bar{h}_{{\rm echo}})}}.
 \label{eq:define_EFA}
\end{equation}
This value roughly estimates the amplitude of the
echo signal when we project the data by using the 
template $\bar{h}_{\rm echo}$ relative to the amplitude of $h^\infty$. 
In Table~\ref{tab:overlap} we show the values of $\rho$ and
EFA for each model with
$a=0.7$, $L=0.9L_c^+$ and $r_{*b}=-100M$.
When $\omega_c$ in the cutoff model gets large,
the overlap decreases because the fraction of the
high frequency modes in $h_{{\rm echo}}$, which is not
included in $\bar{h}_{{\rm echo}}$, increases.
On the other hand, the EFA gets smaller with the decrease of
$\omega_c$ because setting $\omega_c$ smaller simply reduce 
the signal contained in the data. 
A similar behavior of $\rho$ and EFA can be found in the Gaussian
models.
The EFA for the QBH model is very small compared to the
other cases because the amplitude of the echoes is largely 
suppressed by the transfer function as shown in the right panel of 
Fig.~\ref{fig:transfer-fun}.

\begin{table}[t]
\centering
\begin{tabular}{ccccccc}
\hline
model & $\rho$ & EFA & $\rho'$ & $\mbox{EFA}'$
      & $\gamma$ \\
\hline\hline
(cutoff) & & & & & \\
$\omega_c=0.6$ & $0.903$ & $0.347$ & $0.291$
 & $0.112$ & $0.72$ \\
 & $(0.895)$ & $(0.217)$ & $(0.140)$ & $(0.0340)$
 & $(0.51)$ \\
$\omega_c=0.8$ & $0.673$ & $0.588$ & $0.162$
 & $0.142$ & $0.62$ \\
 & $(0.727)$ & $(0.265)$ & $(0.105)$ & $(0.0385)$
 & $(0.52)$ \\
$\omega_c=1$ & $0.443$ & $0.623$ & $0.101$
 & $0.142$ & $0.63$ \\
 & $(0.618)$ & $(0.270)$ & $(0.0885)$ & $(0.0386)$
 & $(0.51)$ \\
$\omega_c=2$ & $0.144$ & $0.627$ & $0.0327$
 & $0.142$ & $0.63$ \\
 & $(0.493)$ & $(0.270)$ & $(0.0705)$ & $(0.0386)$
 & $(0.51)$ \\ \hline
(Gaussian) & & & & & \\
$\sigma_\omega=0.1$ & $0.945$ & $0.329$ & $0.303$
 & $0.105$ & $0.60$ \\
 & $(0.951)$ & $(0.150)$ & $(0.129)$ & $(0.0203)$
 & $(0.53)$ \\
$\sigma_\omega=0.2$ & $0.839$ & $0.445$ & $0.245$
 & $0.130$ & $0.63$ \\
 & $(0.884)$ & $(0.198)$ & $(0.111)$ & $(0.0249)$
 & $(0.50)$ \\
$\sigma_\omega=0.3$ & $0.724$ & $0.508$ & $0.195$
 & $0.137$ & $0.63$ \\
 & $(0.824)$ & $(0.223)$ & $(0.0977)$ & $(0.0264)$
 & $(0.49)$ \\ \hline
QBH & $0.962$ & $0.00554$ & $0.537$
 & $0.00309$ & $0.13$ \\
 & $(0.944)$ & $(0.00263)$ & $(0.248)$
 & $(0.000691)$ & $(0.12)$ \\
\hline
\end{tabular}
\caption{Overlap and EFA between two waveforms of echoes
for $(l,m)=(2,2)$.
The values are computed for $(a,L)=(0.7, 0.9L_c^+)$ and
$r_{*b}=-100M$. The corresponding values for $L=0.1L_c^+$
are shown in parentheses for comparison.}
\label{tab:overlap}
\end{table}

Next we evaluate the decline rates of the echoes.
The echo term
${{\cal K}}Z^H$ in Eq.~(\ref{eq:hatZ8_Z8H}) is composed of 
the sum of the contributions from individual echoes, which correspond 
to the respective terms in the series expansion of the
transfer function,  
\begin{equation}
{{\cal K}} Z^H =
\sum_{n=1}^\infty {{\cal K}}^{(n)} Z^H, \quad
{{\cal K}}^{(n)} \equiv ({\rm R}_b \Phi \hat{r})^{n-1}
 ({\rm R}_b \Phi \hat{t}) \frac{\Gamma^{\infty,+}}{\Gamma^{H,-}}\,.
\end{equation}
Namely, ${{\cal K}}^{(n)} Z^H$ corresponds to the amplitude of the $n$-th echo.
From this, the whole echo waveform is expressed by a simple additive 
sum of the waveforms of the individual echoes as
\begin{equation}
h_{{\rm echo}} = \sum_{n=1}^\infty h_{{\rm echo}}^{(n)}, \quad
h_{{\rm echo}}^{(n)} =
\int d\omega e^{-i\omega u} H(\omega; {{\cal K}}^{(n)} Z^H)
\label{eq:waveform-individual-echoes}
\end{equation}

\begin{figure}[!h]
\centering\includegraphics[width=7cm]{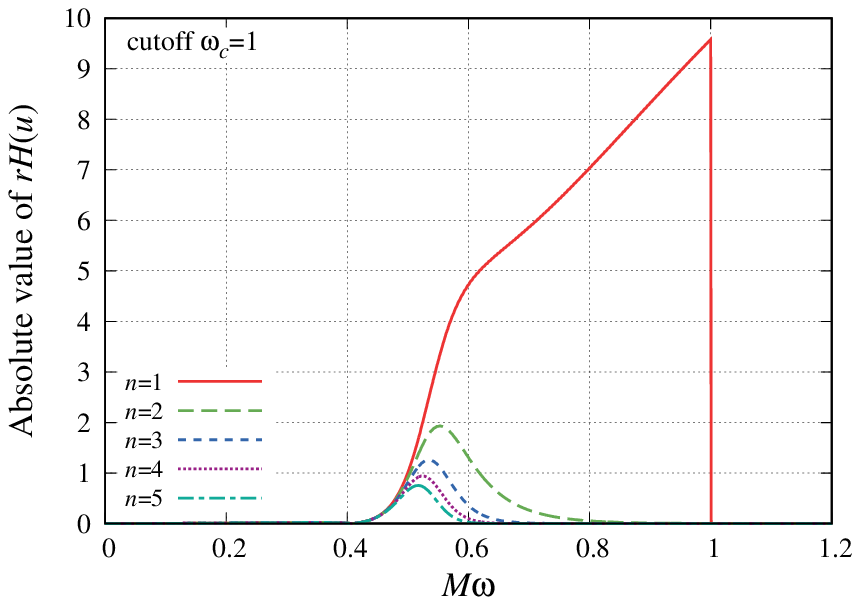}
\centering\includegraphics[width=7cm]{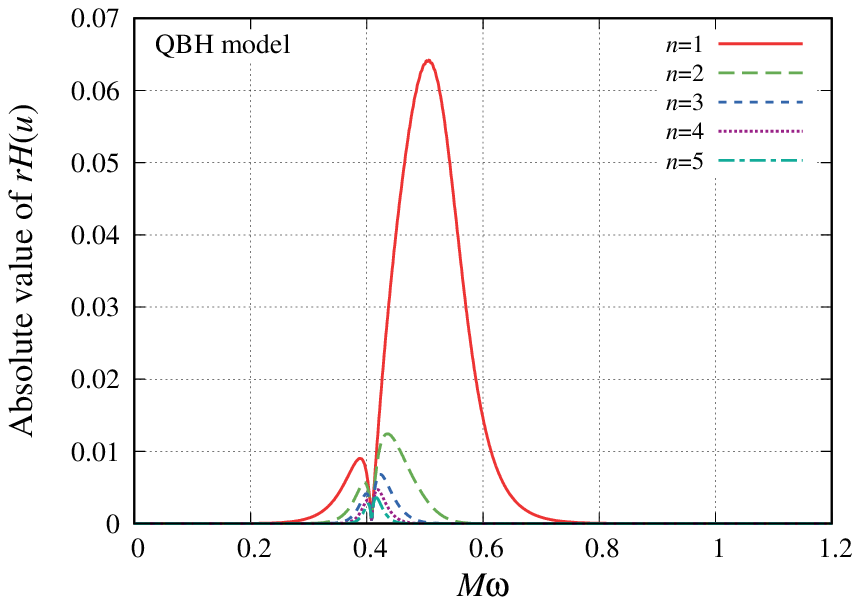}
\caption{Plots of the absolute values of $H(\omega;{\cal K}^{(n)}Z^H)$
for $(l,m)=(2,2)$. 
The left panel is the plot for the cutoff model~(\ref{eq:Rb-cutoff}) with $\omega_c=1$,
while the right panel for the QBH model~(\ref{eq:Rb-Q-model}).}
\label{fig:amplitude_KnHH}
\end{figure}

In Fig.~\ref{fig:amplitude_KnHH} we show 
$|H(\omega;{\cal K}^{(n)}Z^H)|$, 
the absolute values of the waveform of the $n$-th echo for 
two representative models. 
The left panel is the plot for the cutoff model given in
Eq.~(\ref{eq:Rb-cutoff}) with $\omega_c=1$, while 
the right panel for the QBH model given in Eq.~(\ref{eq:Rb-Q-model}). 
The cutoff model is identical to the case of the perfectly 
reflective boundary for $M\omega<1$. The first echo contains 
a large amplitude of high frequency modes, but they  
disappear in the second and later echoes since the reflectance 
due to the angular momentum barrier, $\hat{r}$, is almost zero. 
We also present the case of QBH, because the relative amplitude below 
the threshold frequency of the super-radiance instability is 
significantly larger in the late-time echoes than the other cases. 
Of course, this is not because the lower frequency modes are 
enhanced but because the higher frequency modes are largely 
suppressed in the QBH model. 

\begin{figure}[!h]
\centering\includegraphics[width=7cm]{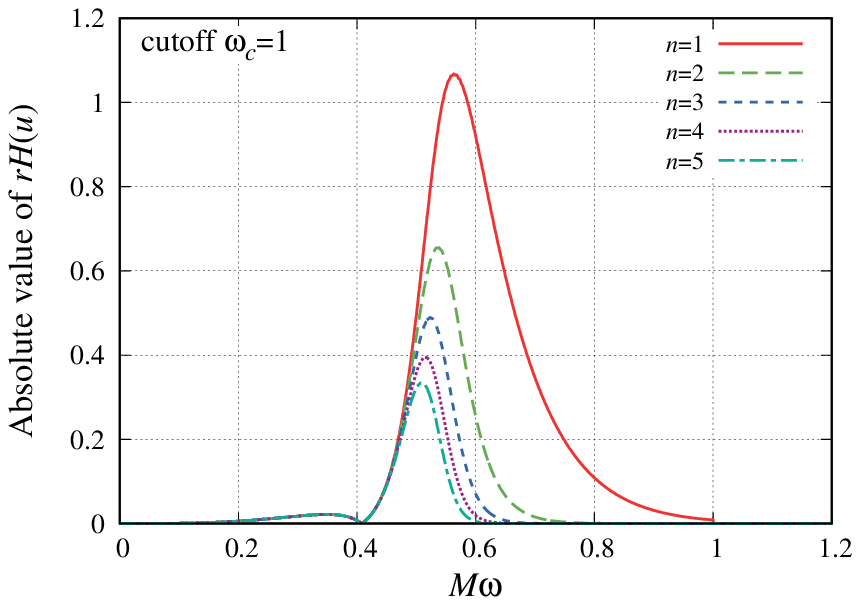}
\centering\includegraphics[width=7cm]{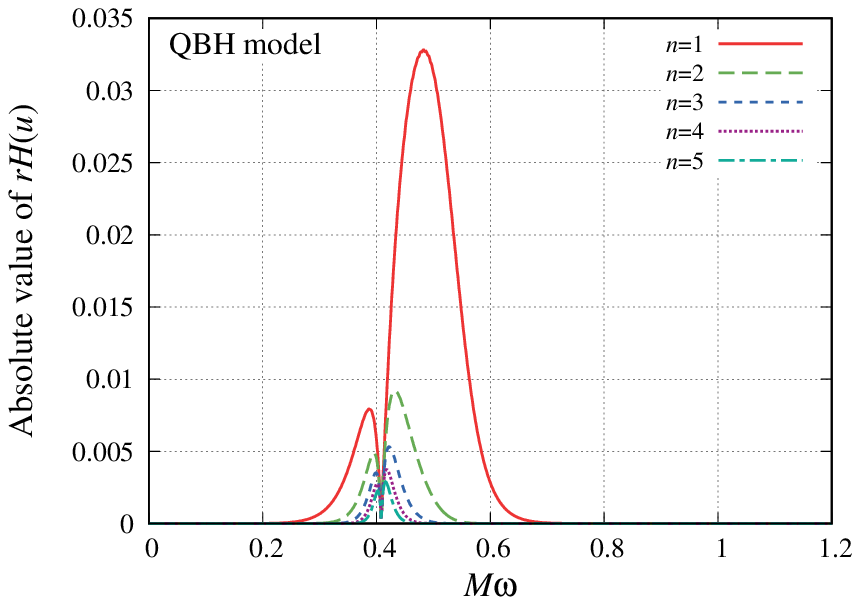}
\caption{Plots of the absolute values of $H(\omega;{\cal K}^{(n)}Z^\infty)$
for $(l,m)=(2,2)$. 
The left panel is for the cutoff model in
Eq.~(\ref{eq:Rb-cutoff}) with $\omega_c=1$,
the right panel for the QBH model in
Eq.~(\ref{eq:Rb-Q-model}).}
\label{fig:amplitude_KnH8}
\end{figure}

In Fig.~\ref{fig:amplitude_KnH8} 
we also plot $|H(\omega;{\cal K}^{(n)}Z^\infty)|$, 
the absolute values of the waveform of 
the $n$-th echo generated by using $Z^\infty$ as the seed, 
instead of $Z^H$. 
The left panel is the plot for the cutoff model with $\omega_c=1$,
while the right panel for the QBH model.

With the identification of the 
waveform of each echo mentioned above, 
we define the following quantities
\begin{equation}
A_n \equiv
\frac{(h_{{\rm echo}}^{(n)}|h_{{\rm echo}}^{(n)})}
{(h_{{\rm echo}}|h_{{\rm echo}})}\,, \quad
B_n \equiv
\frac{(h_{{\rm echo}}^{(n)}|\bar{h}^{(n)}_{{\rm echo}})}
{\sqrt{(h_{{\rm echo}}|h_{{\rm echo}})}
 \sqrt{(\bar{h}_{{\rm echo}}|\bar{h}_{{\rm echo}})}}\,,
\end{equation}
where $B_n$ is evaluated with the same common 
values of $\Delta t$, $\Delta\phi$ and
$\bar{r}_{*b}$ as those used in the maximization 
in Eq.~(\ref{eq:define_rho}). 
$A_n$ is the relative amplitude of each echo to the total
echoes, and $B_n$ is the relative overlap between the $n$-th
echo component of $h_{{\rm echo}}$ and that of $\bar{h}_{{\rm echo}}$.

\begin{figure}[!ht]
\centering\includegraphics[width=7cm]{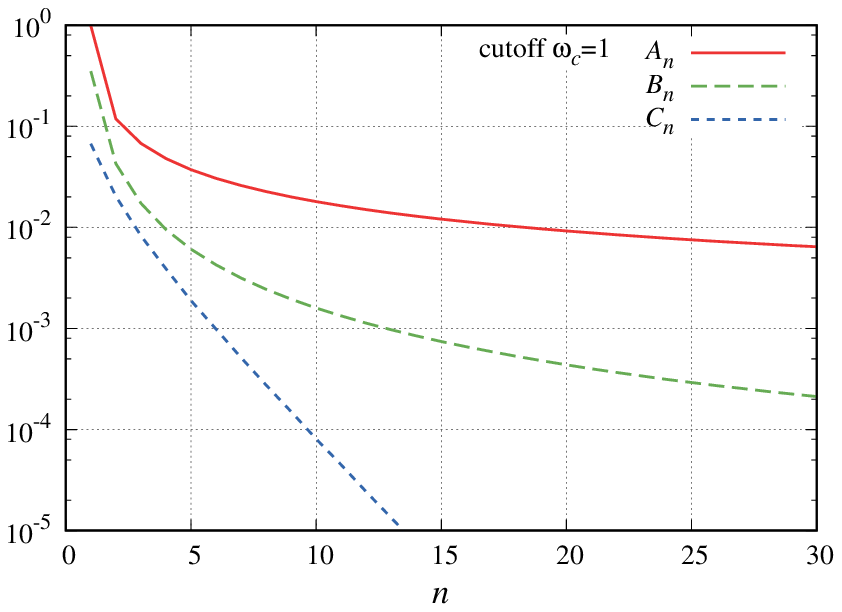}
\centering\includegraphics[width=7cm]{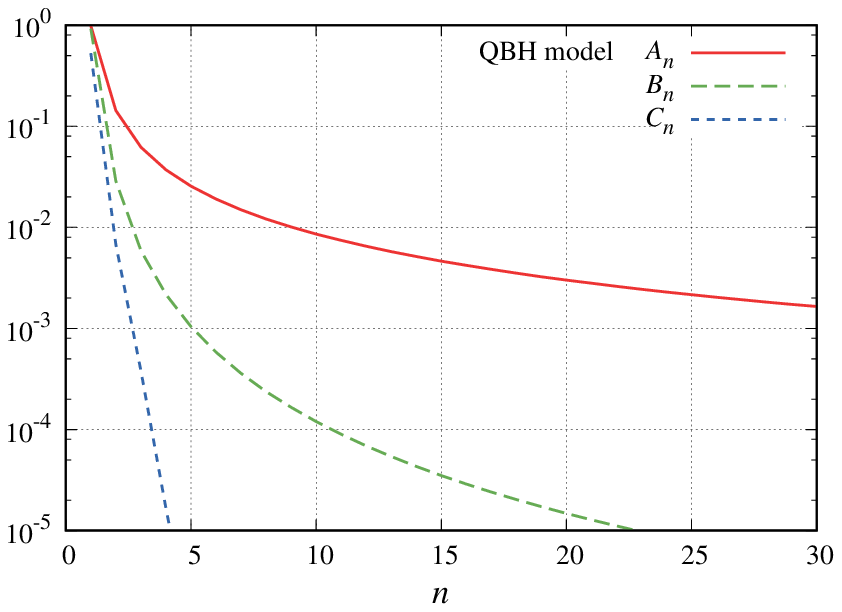}
\caption{Plots of $A_n$, $B_n$ and $C_n$ for $(l,m)=(2,2)$.
The left panel is the plot for the cutoff model~(\ref{eq:Rb-cutoff}) with $\omega_c=1$, and the right panel
for the QNM model~(\ref{eq:Rb-Q-model}).
Here we fix $(a, L)=(0.7, 0.9L_c^+)$ and $r_{*b}=-100M$.}
\label{fig:ABC_vs_n}
\end{figure}

\begin{figure}[!ht]
\centering\includegraphics[width=7cm]{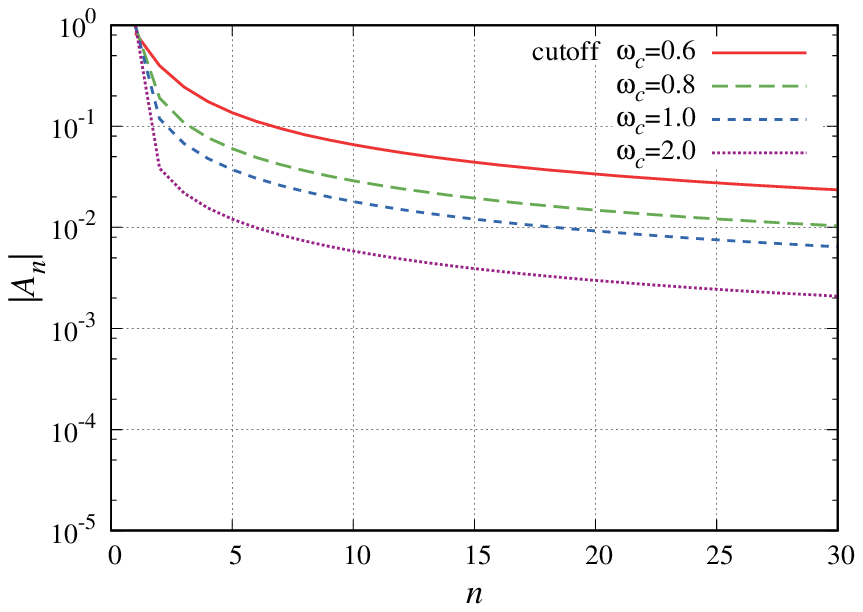}
\centering\includegraphics[width=7cm]{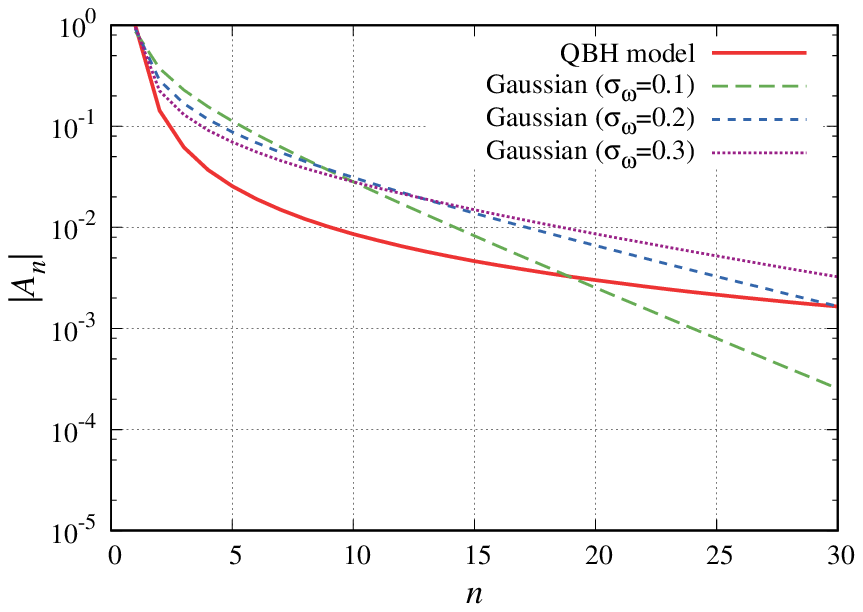} \\
\centering\includegraphics[width=7cm]{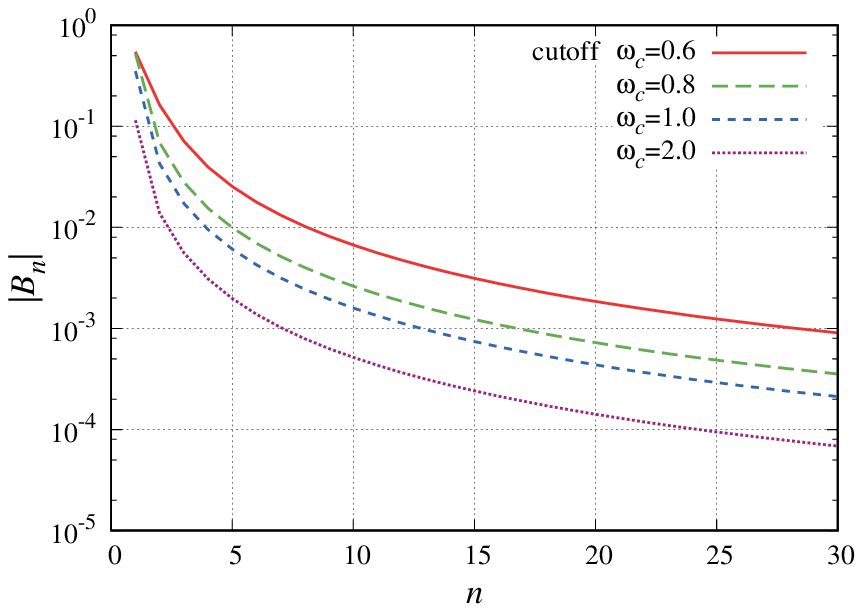}
\centering\includegraphics[width=7cm]{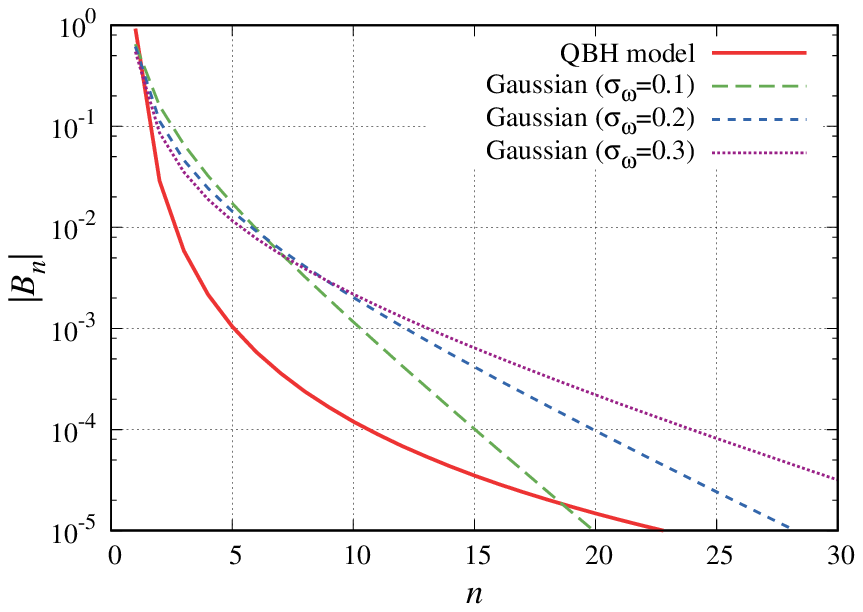} \\
\centering\includegraphics[width=7cm]{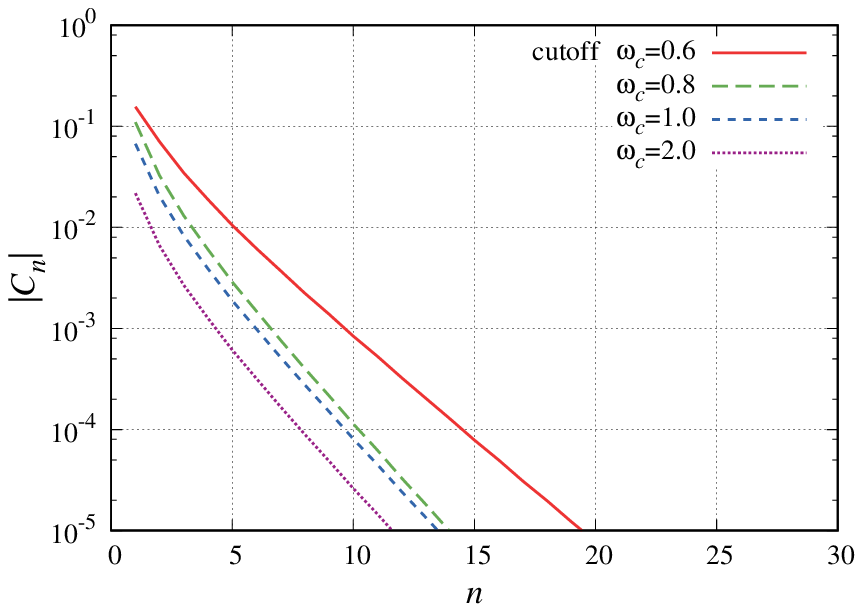}
\centering\includegraphics[width=7cm]{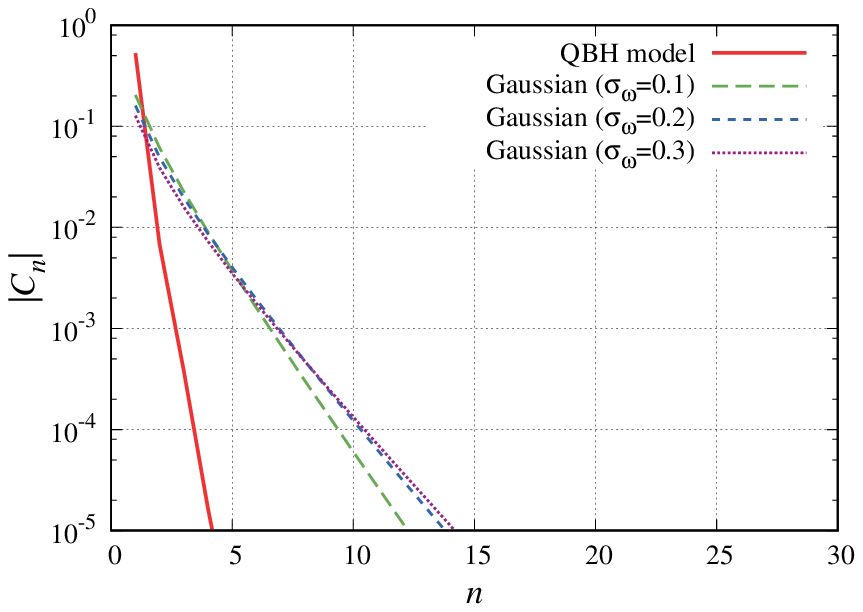}
\caption{Comparison of $A_n$, $B_n$ and $C_n$
for several models. The left panels are the plots 
for the cutoff models while the right panels for the Gaussian models 
and the QBH model.
Here we fix $(a, L)=(0.7, 0.9L_c^+)$ and $r_{*b}=-100M$.}
\label{fig:ABC_vs_n_comparison}
\end{figure}

For comparison, 
we also consider an even simpler waveform model of echoes:
\begin{equation}
\bar{h}'_{{\rm echo}}(u) =
 \sum_{n=1}^{\infty} \bar{h}'^{(n)}_{{\rm echo}}(u),
\quad
\bar{h}'^{(n)}_{{\rm echo}}(u) =
(-1)^n \gamma^{n-1} h^{\infty}(u-(n-1)\Delta t_{{\rm echo}}),
\end{equation}
where $\gamma$ and $\Delta t_{{\rm echo}}$,
independent of the frequency, are corresponding to the
damping factor and the time interval between successive
echoes used in the analysis in Ref.~\cite{Abedi:2016hgu}. 
This model just repeats exactly the same waveform with a minus sign, 
periodically with decaying amplitude. 
We define the overlap and EFA between $h_{{\rm echo}}$ and
$\bar{h}'_{{\rm echo}}$ as
\begin{eqnarray}
\rho' &=&
\max_{\Delta t', \Delta\phi', \gamma, \Delta t_{{\rm echo}}}
\frac{(h_{{\rm echo}}|\bar{h}'_{{\rm echo}})}
{\sqrt{(h_{{\rm echo}}|h_{{\rm echo}})}
 \sqrt{(\bar{h}'_{{\rm echo}}|\bar{h}'_{{\rm echo}})}}\,,
 \label{eq:define_rho-dash} \\
{\rm EFA}' &=&
\max_{\Delta t', \Delta\phi', \gamma, \Delta t_{{\rm echo}}}
\frac{(h_{{\rm echo}}|\bar{h}'_{{\rm echo}})}
{\sqrt{(h^\infty|h^\infty)}
 \sqrt{(\bar{h}'_{{\rm echo}}|\bar{h}'_{{\rm echo}})}}\,,
 \label{eq:define_EFA-dash}
\end{eqnarray}
where we marginalize $\Delta t'$, $\Delta\phi'$, $\gamma$
and $\Delta t_{{\rm echo}}$ to maximize $\rho'$ and ${\rm EFA}'$.
In Table~\ref{tab:overlap}, we show the values of $\rho'$, ${\rm EFA}'$ and $\gamma$ for the maximization. Both $\rho'$ and ${\rm  EFA}'$ are significantly 
smaller than $\rho$ and EFA. This means that the template 
proposed in Ref.~\cite{Nakano:2017fvh} better captures the feature 
of the echo signal expected by the model with a reflective boundary 
near the horizon. At the same time we find that the best fit 
value for the decay rate $\gamma$ is not so large. This is 
not consistent with the results of the data analysis 
presented in Ref.~\cite{Abedi:2016hgu}.  

In a similar manner to $B_n$, we also define
\begin{equation}
C_n \equiv
\frac{(h_{{\rm echo}}^{(n)}|\bar{h}'^{(n)}_{{\rm echo}})}
{\sqrt{(h_{{\rm echo}}|h_{{\rm echo}})}
 \sqrt{(\bar{h}'_{{\rm echo}}|\bar{h}'_{{\rm echo}})}}\,.
\end{equation}
In Figs.~\ref{fig:ABC_vs_n} and \ref{fig:ABC_vs_n_comparison}, 
we show the plots of $A_n$, $B_n$ and $C_n$ as functions of
$n$ for a representative case with $a=0.7$ and $L=0.9L_c^+$.
$B_n$ is much smaller than $A_n$, which means that the echo 
waveform by using GWs emitted to infinity as the seed is 
not really a good approximation. Nevertheless, $B_n$ 
decreases much less rapidly than $C_n$.  

In Fig.~\ref{fig:ABC_vs_n_comparison}, we give a 
comparison of $A_n$, $B_n$ and $C_n$ for several models. 
Here, we adopt $(a, L)=(0.7, 0.9L_c^+)$ as the representative values.
The left panels are the plots 
for the cutoff models. 
All the cases behave very similarly, {\it i.e.}, 
the higher cutoff frequency leads 
to the smaller magnitude for the second and later echoes 
because the higher frequency modes, 
which are contained only in the first echo waveform, are emphasized. 
In the right panels we give the plots 
for the Gaussian models and the QBH model.
These plots show that 
the Gaussian model with broader frequency band decays more rapidly 
at the beginning but more slowly at a late time. 
This is because the model 
contains both the rapidly escaping high frequency modes and 
the long-lasting low frequency modes. 

\section{Conclusion}
We have investigated the expected feature of the waveform of 
GW echoes in the model with a hypothetical reflective boundary 
near the horizon. 
As a model which is easy to handle, we consider perturbations induced 
by a particle falling into a Kerr black hole, instead of a binary 
coalescence. In the latter case, it is not so clear how to impose 
the modified reflective boundary condition 
near the horizon of the black hole newly formed after the merger. 
By contrast, 
imposing the reflective boundary condition at the boundary is 
mathematically well-posed in the case of black hole perturbation. 

We used the Sasaki-Nakamura equation to calculate GWs induced by a 
point mass falling into a Kerr black hole. 
We clarified the method how to compute GWs 
absorbed by the black hole by using the Sasaki-Nakamura equation, 
and developed a prescription to impose the 
reflective boundary condition near the horizon. 
For simplicity, our computation was restricted to the case in  
which the point mass is in an equatorial orbit and initially 
at rest at $r=\infty$, but its angular momentum was varied. 
Independently of 
whether the angular momentum is small or large, the obtained 
spectrum of GWs absorbed by the black hole is dominated by 
modes with a higher frequency than that of 
the fundamental quasi-normal mode.
As a result, the echo signal obtained by introducing 
a reflective boundary near the horizon is also dominated by 
high frequency modes. 

If we assume that the echo waveform is given by a 
simple repetition of the waveform of GWs emitted to infinity, 
a significantly large fraction of echo signal is contained in the 
frequency band lower than the QNM frequency. 
However, our analysis suggests that a simple reflective boundary model 
will not predict large power in echoes at such low frequencies. 

In Ref.~\cite{Uchikata:2019frs} we reanalyzed LIGO data searching 
for the echo signal after binary black hole merger. 
However, the use of our templates that take into account the 
reflection rate of the angular momentum barrier 
did not improve the significance of the signal suggested in Ref.~\cite{Abedi:2016hgu}. The main difference in the templates 
used in these two analyses is in the low frequency bands. 
If there exists echo signal dominated by lower frequency modes, 
our analysis presented in this paper suggests that 
we need to consider more complicated model than the model 
with a simple reflective boundary near the horizon. 

\section*{Acknowledgment}
This work was supported by JSPS KAKENHI Grant Number JP17H06358 (and also JP17H06357), A01: {\it Testing gravity theories using GWs}, as a part of the innovative research area, ``GW physics and astronomy: Genesis''. 
T.T. also acknowledges the support from JSPS KAKENHI Grant No. JP20K03928.
The work of N.S. is partly supported by  JSPS Grant-in-Aid for Scientific Research (C), No. JP16K05356, and
Osaka City University Advanced Mathematical Institute (MEXT Joint Usage/Research Center on Mathematics and Theoretical Physics JPMXP0619217849). Some numerical computations were carried out at the Yukawa Institute Computer Facility.

\end{document}